\documentclass[pre,amsmath,amssymb,twocolumn,nofootinbib,floatfix]{revtex4-1}
\usepackage[colorlinks=true,linkcolor=blue,urlcolor=blue,citecolor=blue]{hyperref}
\usepackage{graphicx}
\usepackage{times}
\usepackage{color}

\definecolor{orange}{rgb}{1,0.6, 0}
\definecolor{darkergreen}{rgb}{0,0.7,0}
\definecolor{grey}{rgb}{.4,.4,0.4}

\newlength{\bilderlength}
\newcommand{\bilderscale}{0.35}

\newcommand{\usebilderscale}{\bilderscale}
\newcommand{\bilderskip}{\hspace*{0.8ex}}

\newcommand{\diagram}[1]{\settowidth{\bilderlength}{\bilderskip\includegraphics[scale=\usebilderscale]{./#1}\bilderskip}\parbox{\bilderlength}{\bilderskip\includegraphics[scale=\usebilderscale]{./#1}\bilderskip}}

\renewcommand{\doi}[2]{\href{http://dx.doi.org/#1}{#2}}
\newcommand{\arxiv}[1]{\href{http://arxiv.org/abs/#1}{#1}}
\newcommand{\link}[2]{\href{http://#1}{#2}}

\newcommand{\Eq}[1]{Eq.~(\ref{#1})}
\newcommand{\Eqs}[1]{Eqs.~(\ref{#1})}
\newcommand{\eq}[1]{(\ref{#1})}

\newcommand{\bea}{\begin{eqnarray}}
\newcommand{\eea}{\end{eqnarray}}

\newcommand{\rme}{\mathrm{e}}

\newcommand{\nn}{\nonumber}
\renewcommand{\epsilon}{\varepsilon}
\newcommand{\ca}[1]{{\cal #1}}
\newcommand{\be}{\begin{equation}}
\newcommand{\ee}{\end{equation}}
\graphicspath{{figures/},{}}
\newcommand{\Fig}[1]{\includegraphics[width=\columnwidth]{./#1}} 
\newcommand{\pfig}[2]{\includegraphics[width=#1]{./#2}}

\usepackage{tikz}
\usetikzlibrary{arrows,shapes,positioning,decorations.pathmorphing}
\usetikzlibrary{decorations.markings}
\usetikzlibrary{snakes}
\tikzset{snake it/.style={decorate, decoration=snake,segment length=3mm}}
\tikzstyle arrowstyle=[scale=1]
\tikzstyle directed=[postaction={decorate,decoration={markings,
    mark=at position .65 with {\arrow[arrowstyle]{stealth}}}}]
\tikzstyle endreversedirected=[postaction={decorate,decoration={markings,
    mark=at position 1.0 with {\arrow[arrowstyle]{stealth}}}}]
\tikzstyle enddirected=[postaction={decorate,decoration={markings,
    mark=at position 1.0 with {\arrow[arrowstyle]{stealth}}}}]
\tikzstyle reverse directed=[postaction={decorate,decoration={markings,
    mark=at position .65 with {\arrowreversed[arrowstyle]{stealth};}}}]
\usetikzlibrary{decorations.markings}
\tikzset{->-/.style={decoration={
  markings,
  mark=at position #1 with {\arrow{>}}},postaction={decorate}}}

\graphicspath{{figures/},{}}

\newcommand{\PK}{\cite{LeDoussalWiese2002}}
\newcommand{\us}{\cite{MukerjeeBonachelaMunozWiese2022}}

\begin{document}

\title{Depinning in the quenched Kardar-Parisi-Zhang class II: Field theory}
\author{Gauthier Mukerjee, Kay J\"org Wiese}
\affiliation{Laboratoire de Physique de l'\'Ecole Normale Sup\'erieure, ENS, Universit\'e PSL, CNRS, Sorbonne Universit\'e, Universit\'e Paris-Diderot, Sorbonne Paris Cit\'e, 24 rue Lhomond, 75005 Paris, France}  

\begin{abstract}
There are two main universality classes for depinning of elastic interfaces in disordered media: quenched Edwards-Wilkinson (qEW), and quenched Kardar-Parisi-Zhang (qKPZ). The first class is relevant as long as the elastic force between two neighboring sites on the interface is purely harmonic, and invariant under tilting. The second class applies when the elasticity is non-linear, or the surface grows preferentially in its normal direction. It encompasses fluid imbibition, the Tang-Leschorn cellular automaton of 1992 (TL92), depinning with anharmonic elasticity (aDep), and qKPZ. 
While the field theory is well developed for qEW, there is no consistent theory for qKPZ. The aim of this paper is to  construct this field theory  within the Functional renormalization group (FRG) framework, based   on large-scale numerical simulations in dimensions $d=1$, $2$ and $3$, presented in a companion paper. In order to measure the effective force correlator and coupling constants, the driving force is derived from a confining potential with curvature $m^2$. 
We show, that contrary to common belief this is allowed in the presence of a KPZ term. The ensuing field theory becomes massive, and can no longer be  Cole-Hopf transformed. In exchange, it possesses an IR attractive stable fixed point at a finite KPZ non-linearity $\lambda$. 
Since there is neither elasticity nor a KPZ term in dimension $d=0$, qEW and qKPZ merge there.  As a result, the two universality classes are distinguished by terms linear in $d$. This allows us to build a consistent field theory in dimension $d=1$, which loses some of its predictive powers in higher dimensions.

\end{abstract} 

\maketitle

\section{Introduction} 
\begin{figure}[b]
\Fig{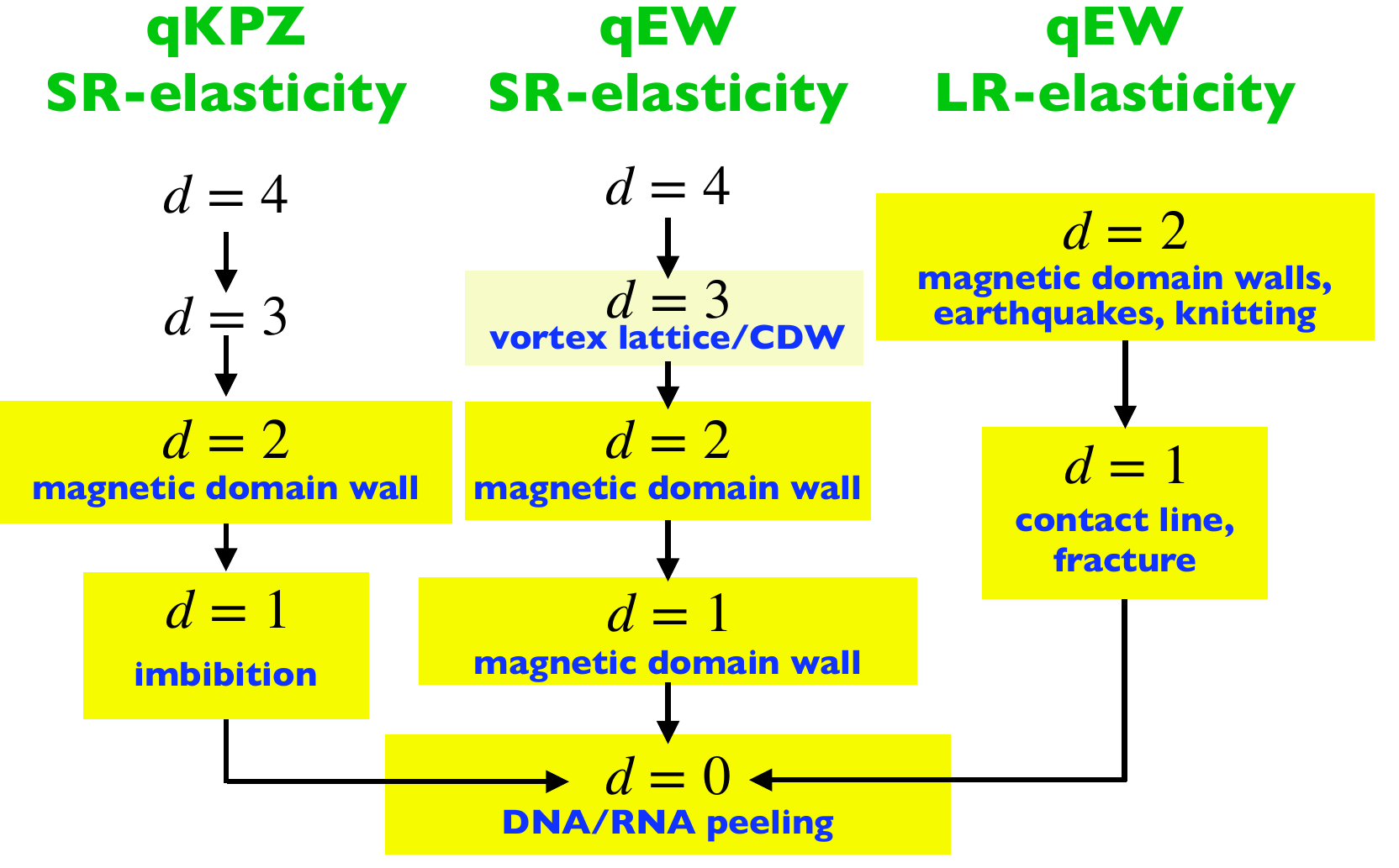}
\caption{Universality classes at depinning, for $d<d_{\rm c}$. For the yellow shaded cases experiments exist.}
\label{f:classes}
\end{figure}

Disordered elastic manifolds exhibit universal critical behavior when driven slowly, known as {\em depinning}.   There are two main universality classes, each associated with a stochastic differential equation of evolution: quenched Edwards-Wilkinson (qEW) \cite{Wiese2021}, and quenched Kardar-Parisi-Zhang (qKPZ) \cite{TangKardarDhar1995}. The first class is relevant as long as the elasticity of the interface is purely harmonic, and invariant under tilting. 
This description is valid in a variety of situations such as magnetic domain walls in the presence of disorder a.k.a.\ the Barkhausen effect \cite{Barkhausen1919,DurinZapperi2006b,DurinBohnCorreaSommerDoussalWiese2016,terBurgBohnDurinSommerWiese2021}, vortex 
lattices \cite{ScheidlVinokur1998,BucheliWagnerGeshkenbeinLarkinBlatter1998},   charge-density waves \cite{NarayanDSFisher1992b}, and DNA unzipping \cite{WieseBercyMelkonyanBizebard2019}. While these systems have short-ranged elasticity, this framework can readily be adapted to describe systems with long-range (LR) elasticity such as 
contact-line depinning \cite{LeDoussalWieseRaphaelGolestanian2004,BachasLeDoussalWiese2006,LeDoussalWieseMoulinetRolley2009},  earthquakes \cite{DSFisher1998,FisherDahmenRamanathanBenZion1997} and knitting \cite{PoinclouxAdda-BediaLechenault2018b,PoinclouxAdda-BediaLechenault2018}.

The second class is relevant when the elasticity is non-linear, or the surface grows preferentially in its normal direction. It encompasses fluid imbibition \cite{BuldyrevBarabasiCasertaHavlinStanleyVicsek1992}, the Tang-Leschorn cellular automaton of 1992 (TL92) \cite{TangLeschhorn1992} or its variants  \cite{AmaralBarabasiBuldyrevHarringtonHavlinSadr-LahijanyStanley1995}, depinning with anharmonic elasticity (aDep) \cite{RossoKrauth2001b}, and qKPZ \cite{TangKardarDhar1995}. 
That all these models are in the same universality class is non-trivial, but is now firmly established \us.
This so-called qKPZ class has been observed for magnetic domain walls 
\cite{MoonKimYooChoHwangKahngMinShinChoe2013,Diaz-PardoMoisanAlbornozLemaitreCurialeJeudy2019}, in growing bacterial colonies  \cite{HuergoMuzzioPasqualeGonzalezBolzanArvia2014}  and chemical reaction fronts \cite{AtisDubeySalinTalonLeDoussalWiese2014}.

While the field theory for qEW is well established \cite{NarayanDSFisher1992a,NarayanDSFisher1993a,LeschhornNattermannStepanowTang1997,NattermannStepanowTangLeschhorn1992,ChauveLeDoussalWiese2000a,LeDoussalWieseChauve2002,LeDoussalWieseChauve2003,Wiese2021}, 
building a field theory for qKPZ is a challenge. 
It has previously been attempted in Ref.~\cite{LeDoussalWiese2002}. In that work, the  running coupling constant for the non-linearity goes  to infinity. 
The first question one needed to clarify was whether this is true, or an artifact of the Functional Renormalization Group (FRG) treatment. 
In Ref.~\cite{MukerjeeBonachelaMunozWiese2022} we measured in a numerical simulation the effective action of three models: qKPZ, TL92, and aDep. For $d=1$ we found that all three possess an effective long-distance behavior fully described  by the terms in the qKPZ equation, 
\bea
\eta \partial_t u(x,t) &=& c\nabla^2 u(x,t) +\lambda \left[\nabla u(x,t)  \right]^2+
m^2\big[w{-} u(x,t)\big] \nn\\
&& + F\big(x,u(x,t)\big).
\label{eq:qKPZ}
\eea
The disorder forces $F(x,u)$ are  quenched Gaussian random variables with 
variance 
\be\label{F-var}
\overline{F(x,u)F(x',u')}= \delta^d(x-x') \Delta_0(u-u').
\ee $\Delta_0(u)$ is the  microscopic disorder-force correlator, assumed to decay rapidly for short-range (SR) disorder.
In higher dimensions the same conclusions were reached, although with   larger uncertainties. 

Motivated by these findings, we reconsider the field theory corresponding to \Eq{eq:qKPZ}. There are two  key observations.
First of all,  for $d\to 0$, three universality classes   merge: qKPZ, qEW with short-ranged, and qEW with LR interactions. This is visualized  on Fig.~\ref{f:classes}. 
The second key observation is that the way  we drive the system is important. In fact, we drive with a force that derives from a confining potential, the term $  m^2[w-u(x,t)]$ in \Eq{eq:qKPZ}. That allows us to measure the  effective force correlator $\Delta(w)$ defined via
\bea\label{Delta-def}
\Delta(w-w') &:=& m^4 L^d\, \overline{ (u_w-w) (u_{w'}-w')}^{\rm c}, \\
u_w &:=& \frac1{L^d} \int_x u_w(x), \label{eq:uw}\\
u_w(x) &:=& \lim _{t\to \infty} u(x,t) \mbox{~given~}w\mbox{~fixed.}
\eea
In this protocol, $w$ is increased in steps. One then waits until the interface stops, which defines $u_w(x)$. Its center-of-mass position is $u_w$, and its fluctuations define $\Delta(w)$.

\begin{figure}[t]
\centerline{\includegraphics[width=1.0\linewidth]{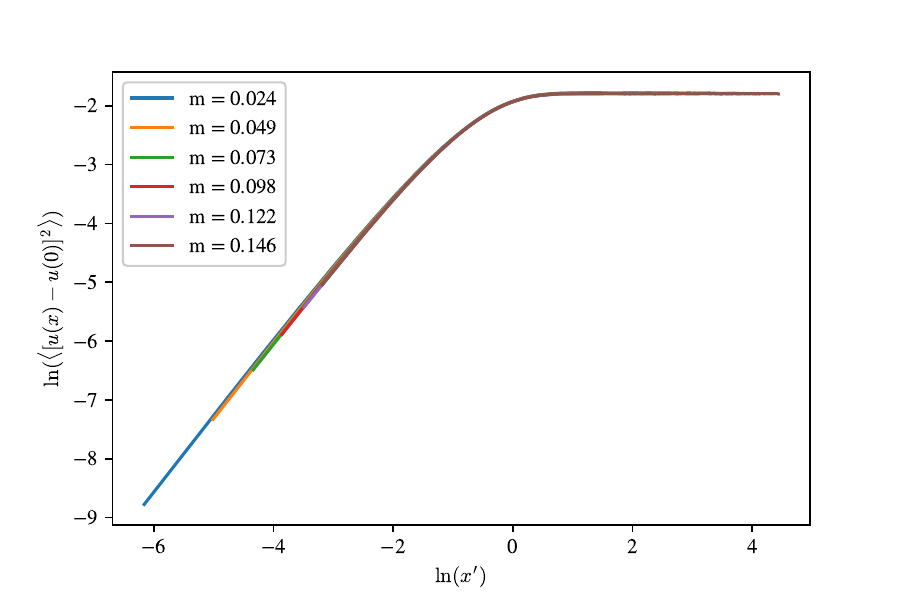}}
\caption{Scaling collapse of the two point   function \eq{2-pt-function} obtained by rescaling $x$ and $y = \langle[u(x)-u(0)]^2\rangle$ such that $ x^\prime = \frac{x}{\xi_m} = x m^{\frac{\zeta_m}{\zeta}}$ and $y' = y m^{2\zeta_m}$, in logarithmic scale, for TL92 in $d=1$.}
\label{fig:2pcollapse}
\end{figure}

This driving force appears in the effective action \Eq{qKPZ-action} as a mass term, as compared to \cite{LeDoussalWiese2002}, which considered a massless theory. 
Their motivation was that a massive term breaks Galilean invariance, and this is something ``you do not want for the KPZ equation''. 
We believe that this is not a problem here, for two reasons: First of all, Galilean invariance is already broken by the quenched disorder  $F(x,u)$, even after disorder averaging. Second, even if the driving  breaks Galilean invariance, this should only affect large-scale properties, but not small-scale ones (small with respect to the correlation length), and especially not critical properties. 

There is a prize to pay for introducing a massive term: one loses the Cole-Hopf transformation, a transformation that allows to map the KPZ equation to a simpler stochastic heat equation with multiplicative noise (see section \ref{s:Cole-Hopf}). This the authors of Ref.~\cite{LeDoussalWiese2002} were not ready to give up, as it complicates perturbation theory. 
As we will see below, it breaks the non-renormalization of $\lambda/c$, allowing us to find a fixed point for the latter. 
As both the massive and the massless scheme give, at least in 1-loop order, the same results close to the upper critical dimension $d_{\rm c}=4$, what we will present below is not a systematic $\epsilon$-expansion. Rather, if we suppose we know the FRG fixed point for qEW, then our scheme allows us to control 
qKPZ perturbatively in $d$, in an expansion around the qEW fixed point. 
While the latter is known analytically in $d=0$ \cite{LeDoussalWiese2008a}, what we use here is the 1-loop fixed point, obtained via the $\epsilon=4-d$ expansion.
The latter is actually   quite good even down to $d=1$: It predicts a roughness exponent of $\zeta=1$, as compared to the best numerical value of 
$\zeta=5/4$  \cite{GrassbergerDharMohanty2016,ShapiraWieseUnpublished}. The FRG correlator is, even quantitatively, rather well approximated by its 1-loop value
\cite{terBurgBohnDurinSommerWiese2021,Wiese2021}. We restrict ourselves here to 1-loop order, which has the benefit of greater transparency.
Preliminary calculations show that 
extension to 2-loop order is straightforward   though    cumbersome. 
The method we present below allows us to compute analytically the different critical exponents as well as the full force correlator, and present quantitative agreement with the numerical simulations.

This paper is organized as follows:
In the next section \ref{sec:scalingtheory} we first define the field theory and review perturbation theory (section \ref{s:Model and Action}).
We then  summarize scaling arguments described in detail in the companion paper \cite{MukerjeeBonachelaMunozWiese2022} (section \ref{s:Scaling and anomalous exponents}).
The effective  force correlator is defined in section \ref{s:The renormalized correlator Delta(w)}, and  the relation to directed percolation  in section \ref{Link to directed percolation,    exponents given in the literature, and other    relations}, followed by a discussion of the effective action measured in simulations, section \ref{sec:simul}.
Section \ref{sec:fieldtheory} is dedicated to the field theory. 
We start with a reminder on the generation of the KPZ term from an anharmonic elasticity
(section \ref{s:Generation of KPZ term from anharmonic elasticity}). All 1-loop contributions are given in section \ref{s:1-loop contributions}, with details relegated to appendix \ref{sec:diagrams}.  Section \ref{s:Flow equations} establishes the flow equations. 
Necessary conditions for their solution are derived in section \ref{s:Necessary conditions for a fixed point, and bounds}, followed by an analytical solution in section \ref{solution-4-flow}, first giving the scheme (section \ref{Scheme}), and then explicit values in $d=1$ to $d=3$ (sections \ref{d=1} to \ref{d=3}). 
Tables summarize our findings in  section \ref{Other properties}.
We comment on the Cole-Hopf transformation (section \ref{s:Cole-Hopf}), and present in layman terms physical insights from our work section \ref{sec:phys_insight}, before concluding in section \ref{s:Conclusions}.

\section{Model and Phenomenology}\label{sec:scalingtheory}
\label{s:Perturbation theory}

\subsection{Model, action and perturbation theory}
\label{s:Model and Action}
The Martin-Siggia-Rose \cite{MSR} action corresponding to \Eq{eq:qKPZ} reads  
\bea
\label{qKPZ-action}
\ca S[u,\tilde u] &=& \int_{x,t}  \tilde u(x,t)\Big\{\eta \partial_t u(x,t)- c\nabla^2 u(x,t)  \\
&& - \lambda \left[\nabla u(x,t)  \right]^2+
m^2\big[ u(x,t){-}w\big] \Big\} \nn\\
&& -\frac12 \int_{x,t,t'}\tilde u(x,t) \Delta_0\big(u(x,t)-u(x,t')\big) \tilde u(x,t'). \nn 
\eea
The field $\tilde u(x,t)$ is an auxiliary field introduced to enforce \Eq{eq:qKPZ} and called the response field. The last term is obtained by averaging $\rme^{\int_{x,t}\tilde u(x,t) F(x,(u(x,t)) }$ over $F(x,u)$, using its variance \eq{F-var}.
Perturbation theory is constructed by expanding around the free theory obtained by setting  $\lambda\to 0$ and $\Delta_0(u)\to 0 $ in \Eq{qKPZ-action}.
The free response function   is the response of the field $u(x+x',t+t')$ to an additional force acting at $x'$, $t'$, 
\bea
R(x,t) &:=& \left< \frac{\delta u(x+x',t+t')}{\delta f(x',t')} \right> \nn\\
&= &\left<  {\delta u(x+x',t+t')}{\tilde u(x',t')} \right>.
\eea
In Fourier space it reads
\bea
R(k,t)&=& \left< \tilde u(k,t) u(-k,t') \right> \nn\\
 &=& \theta (t'>t) \frac1{\eta}\rme^{-(c k^2+m^2 )(t'-t)/\eta}\nn\\
&=& 
{\parbox{1.2cm}{{\begin{tikzpicture}
\coordinate (x1t1) at  (0,0) ; 
\coordinate (x1t2) at  (1,0) ; 
\fill (x1t1) circle (2pt);
\fill (x1t2) circle (2pt);
\draw [directed] (x1t1) -- (x1t2);
\end{tikzpicture}}}}.\qquad
\eea
Graphically this is represented by an arrow from $\tilde u$ to $u$. The  (microscopic)  disorder  is represented by two dots connected by a dashed line, whereas the KPZ vertex is a dot with two incoming lines with bars for the derivatives, and one outgoing one. Examples for diagrams correcting the disorder are given on Fig.~\ref{all-corrections-4-Delta}.
For an introduction into functional perturbation theory we refer to section 3 of \cite{Wiese2021}.
Note that the disorder is corrected by the KPZ force, and what we   loosely call the {\em renormalized disorder} is more precisely the {\em renormalized force correlator}, which contains contributions from the KPZ term (see Section \ref{s:The renormalized correlator Delta(w)} for a detailed discussion).
Non-trivial correlations necessitate at least one ``disorder" vertex $\Delta_0(u)$. As an example, the leading order to the equal-time 2-point function is
\bea
\setlength{\unitlength}{0.8mm}
\left<   u(k,0) u(-k,0) \right> &=&
{\parbox{2.1cm}{{\begin{tikzpicture}
\coordinate (x1t1) at  (0,0) ; 
\coordinate (x1t2) at  (0,.5) ; 
\coordinate (x2t3) at  (1.5,0) ; 
\coordinate (x2t4) at  (1.5,0.5) ; 
\coordinate (x) at  (0,0)  ; \coordinate (y) at  (1.5,0) ; \fill (x1t1) circle (2pt);
\fill (x1t2) circle (2pt);
\fill (x2t3) circle (2pt);
\fill (x2t4) circle (2pt);
\draw [directed] (x1t1) -- (x2t3);
\draw [directed] (x1t2) -- (x2t4);
\draw [dashed,thick] (x1t1) -- (x1t2);
\end{tikzpicture}}}}
\nn\\
&=&
\left[  \int_{t} R(k,t) \right]^2\Delta_0(0) \nn\\
&  =& \frac{\Delta_0(0)}{(c k^2+m^2)^2}.
\eea
The arrows represent the response function $R$, the dotted line   the  effective force correlator $\Delta_0(u)$.  
We assume an upper critical dimension of $d_{\rm c} = 4$ as in qEW. 
Simulations  show that 
in $d=3$ the interface is still rough \us, so the upper critical dimension is above $3$. Noting that physical realizations can only be constructed in 
integer dimensions, the remaining open question is whether   the 4-dimensional system is at its upper critical dimension, or potentially above,  see  section \ref{d=3}. Note that an interface in anharmonic depinning is less rough than in qEW; this excludes an   upper critical dimension larger than four.

\subsection{Scaling and anomalous exponents}
\label{s:Scaling and anomalous exponents}
Scaling arguments were given in the companion paper \cite{MukerjeeBonachelaMunozWiese2022}. 
We recall the main results here. 
The static  2-point function is defined as 
\be\label{2-pt-function}
\frac{1}{2}\overline{\langle[u(x)-u(y)]^{2}} \simeq \left\{\begin{array}{c}
A|x-y|^{2 \zeta},~|x-y| \ll \xi_{m}, \\
B m^{-2 \zeta_{m}}, ~~~~|x-y| \gg \xi_{m}.
\end{array}\right.
\ee
The average is taken over different disorder configurations (there are no thermal fluctuaions). $\zeta$
is the standard roughness exponent. In contrast to qEW, there is a new exponent $\zeta_m> \zeta$.
The reason is that the elasticity $c$ renormalizes and thus its anomalous dimension gives rise to another exponent.  The quantity $\xi_m$ in \Eq{2-pt-function} is the
correlation length created by the confining potential.
Every length parallel to the interface scales as $x$ or $\xi_m$, 
whereas  in the perpendicular direction it scales as $u \sim x^{\zeta}\sim \xi_m^{\zeta}$. 
To estimate $\xi_m$,   we take $x= \xi_m$ in \Eq{2-pt-function},   obtaining
$ \xi_m^{2 \zeta}  \sim m^{-2\zeta_m}$. As a consequence 
\be
 \xi_m   \sim m^{-\frac{\zeta_m}{\zeta}}.
\ee
Note that
$ \xi_m \not \sim \frac{1}{m}$ as  for qEW. 
Fig.~\ref{fig:2pcollapse} shows a    scaling collapse of the 2-point function with these scalings.

Define  $\psi_{\lambda}$, $\psi_{c}$ and $\psi_{\eta}$ to be the anomalous   dimensions of $\lambda$, $c$ and $\eta$ in units of $m^{-1}$, 
\bea
\label{Psi-c}
\psi_{\rm c} &:=&-  m \partial_m \ln (c), \\
\label{Psi-lambda}
\psi_\lambda &:=&  - m \partial_m \ln(\lambda),  \\
\label{Psi-eta}
\psi_\eta &:=&   - m \partial_m \ln (\eta )  .
\eea
In order to relate them  to the standard scaling exponents $\zeta$, $\zeta_m$ and $z$, we first need to define $z$. It is given by the temporal spread of the perturbation in the surface  in the 2-point function as
\be
\frac{1}{2}\overline{[u(x,t)-u(x,t')]^{2}} \sim |t-t'|^{2\zeta/z}.
\ee
With these definitions at the fixed point we can derive
\begin{align}
\label{9}
\frac{\zeta_m}{\zeta} &= 1+ \frac {\psi_{\rm c}}2, \\
 \label{10}
 \zeta_m  &=  \psi_{\rm c} - \psi_\lambda, \\
  \label{11}
  z &=  \frac{\zeta}{\zeta_m}(2+\psi_\eta) .
\end{align}
{The first relation is obtained from $x^{-1}\sim q \sim m/\sqrt{c}\sim m^{1+\psi_{\rm c}/2}$, implying $x^{2\zeta}\sim m^{-(2+\psi_{\rm c})\zeta} \equiv m^{-2\zeta_m}$. The second follows from $\lambda u \sim c$. The last one is obtained from $\eta/t \sim m^2$, implying $t\sim m^{-2-\psi_\eta}\sim x^{(2+\psi_\eta)\zeta/\zeta_m}$.}

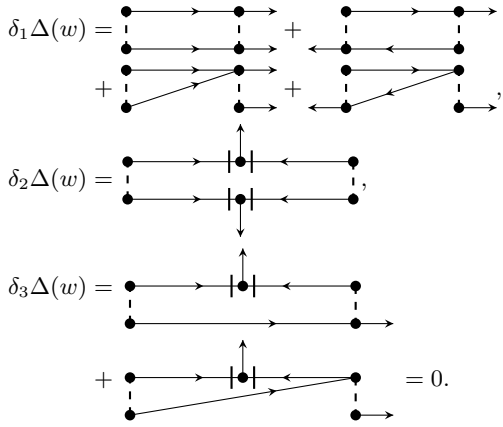
\begin{figure}[t]
\begin{eqnarray*}
\setlength{\unitlength}{0.8mm}
\delta_1 \Delta(w) &=&{\parbox{2.1cm}{{\begin{tikzpicture}
\coordinate (x1t1) at  (0,0) ; 
\coordinate (x1t2) at  (0,.5) ; 
\coordinate (x2t3) at  (1.5,0) ; 
\coordinate (x2t4) at  (1.5,0.5) ; 
\coordinate (x) at  (0,0)  ; \coordinate (y) at  (1.5,0) ; \fill (x1t1) circle (2pt);
\fill (x1t2) circle (2pt);
\fill (x2t3) circle (2pt);
\fill (x2t4) circle (2pt);
\draw [directed] (x1t1) -- (x2t3);
\draw [directed] (x1t2) -- (x2t4);
\draw [dashed,thick] (x1t1) -- (x1t2);
\draw [dashed,thick] (x2t3) -- (x2t4);
\draw [enddirected]  (x2t3)--(2,0);
\draw [enddirected]  (x2t4)--(2,0.5);
\end{tikzpicture}}}} + {\parbox{2.5cm}{{\begin{tikzpicture}
\coordinate (x1t1) at  (0,0) ; 
\coordinate (x1t2) at  (0,.5) ; 
\coordinate (x2t3) at  (1.5,0) ; 
\coordinate (x2t4) at  (1.5,0.5) ; 
\coordinate (x) at  (0,0)  ; \coordinate (y) at  (1.5,0) ; \fill (x1t1) circle (2pt);
\fill (x1t2) circle (2pt);
\fill (x2t3) circle (2pt);
\fill (x2t4) circle (2pt);
\draw [directed]  (x2t3)--(x1t1) ;
\draw [directed] (x1t2) -- (x2t4);
\draw [dashed,thick] (x1t1) -- (x1t2);
\draw [dashed,thick] (x2t3) -- (x2t4);
\draw [enddirected]  (x1t1)--(-.5,0);
\draw [enddirected]  (x2t4)--(2,0.5);
\end{tikzpicture}}}}
\\
 &+& {\parbox{2.1cm}{{\begin{tikzpicture}
\coordinate (x1t1) at  (0,0) ; 
\coordinate (x1t2) at  (0,.5) ; 
\coordinate (x2t3) at  (1.5,0) ; 
\coordinate (x2t4) at  (1.5,0.5) ; 
\coordinate (x) at  (0,0)  ; \coordinate (y) at  (1.5,0) ;  \fill (x1t1) circle (2pt);
\fill (x1t2) circle (2pt);
\fill (x2t3) circle (2pt);
\fill (x2t4) circle (2pt);
\draw [directed] (x1t1) -- (x2t4);
\draw [directed] (x1t2) -- (x2t4);
\draw [dashed,thick] (x1t1) -- (x1t2);
\draw [dashed,thick] (x2t3) -- (x2t4);
\draw [enddirected]  (x2t3)--(2,0);
\draw [enddirected]  (x2t4)--(2,0.5);
\end{tikzpicture}}}} + {\parbox{2.5cm}{{\begin{tikzpicture}
\coordinate (x1t1) at  (0,0) ; 
\coordinate (x1t2) at  (0,.5) ; 
\coordinate (x2t3) at  (1.5,0) ; 
\coordinate (x2t4) at  (1.5,0.5) ; 
\coordinate (x) at  (0,0)  ;\coordinate (y) at  (1.5,0) ;\fill (x1t1) circle (2pt);
\fill (x1t2) circle (2pt);
\fill (x2t3) circle (2pt);
\fill (x2t4) circle (2pt);
\draw [directed]  (x2t4)--(x1t1) ;
\draw [directed] (x1t2) -- (x2t4);
\draw [dashed,thick] (x1t1) -- (x1t2);
\draw [dashed,thick] (x2t3) -- (x2t4);
\draw [enddirected]  (x1t1)--(-.5,0);
\draw [enddirected]  (x2t3)--(2,0.);
\end{tikzpicture}}}},
\\
\delta_2 \Delta (w) &=&
{\parbox{3.2cm}{{\begin{tikzpicture}
\coordinate (x1t1) at  (0,0) ; 
\coordinate (x1t2) at  (0,.5) ; 
\coordinate (x2t3) at  (1.5,0) ; 
\coordinate (x2t4) at  (1.5,0.5) ; 
\coordinate (x) at  (0,0)  ; 
\coordinate (y) at  (1.5,0) ; 
\coordinate (x2) at  (3,0.5)  ; 
\coordinate (y2) at  (3,0) ; 
\fill (x1t1) circle (2pt);
\fill (x1t2) circle (2pt);
\fill (x2t3) circle (2pt);
\fill (x2t4) circle (2pt);
\fill (x2) circle (2pt);
\fill (y2) circle (2pt);
\draw [directed] (x1t1) -- (x2t3);
\draw [directed] (x1t2) -- (x2t4);
\draw [directed] (x2) -- (x2t4);
\draw [directed] (y2)  -- (x2t3);
\draw [dashed,thick] (x1t1) -- (x1t2);
\draw [dashed,thick] (x2) -- (y2);
\draw [enddirected]  (x2t3)--(1.5,-.5);
\draw [enddirected]  (x2t4)--(1.5,1);
\draw [thick] (1.35,-.15) -- (1.35,.15);
\draw [thick] (1.35,.35) -- (1.35,.65);
\draw [thick] (1.65,-.15) -- (1.65,.15);
\draw [thick] (1.65,.35) -- (1.65,.65);
\end{tikzpicture}}}},
\\
\delta_3 \Delta (w)&=&
{\parbox{3.7cm}{{\begin{tikzpicture}
\coordinate (x1t1) at  (0,0) ; 
\coordinate (x1t2) at  (0,.5) ; 
\coordinate (x2t4) at  (1.5,0.5) ; 
\coordinate (x) at  (0,0)  ; 
\coordinate (y) at  (1.5,0) ; 
\coordinate (x2) at  (3,0.5)  ; 
\coordinate (y2) at  (3,0) ; 
\fill (x1t1) circle (2pt);
\fill (x1t2) circle (2pt);
\fill (x2t4) circle (2pt);
\fill (x2) circle (2pt);
\fill (y2) circle (2pt);
\draw [directed] (x1t1) -- (y2);
\draw [directed] (x1t2) -- (x2t4);
\draw [directed] (x2) -- (x2t4);
\draw [dashed,thick] (x1t1) -- (x1t2);
\draw [dashed,thick] (x2) -- (y2);
\draw [enddirected]  (x2t4)--(1.5,1);
\draw [enddirected]  (y2)--(3.5,0.);
\draw [thick] (1.35,.35) -- (1.35,.65);
\draw [thick] (1.65,.35) -- (1.65,.65);
\end{tikzpicture}}}} \nn\\
 &+& 
{\parbox{3.7cm}{{\begin{tikzpicture}
\coordinate (x1t1) at  (0,0) ; 
\coordinate (x1t2) at  (0,.5) ; 
\coordinate (x2t4) at  (1.5,0.5) ; 
\coordinate (x) at  (0,0)  ; 
\coordinate (y) at  (1.5,0) ; 
\coordinate (x2) at  (3,0.5)  ; 
\coordinate (y2) at  (3,0) ; 
\fill (x1t1) circle (2pt);
\fill (x1t2) circle (2pt);
\fill (x2t4) circle (2pt);
\fill (x2) circle (2pt);
\fill (y2) circle (2pt);
\draw [directed] (x1t1) -- (x2);
\draw [directed] (x1t2) -- (x2t4);
\draw [directed] (x2) -- (x2t4);
\draw [dashed,thick] (x1t1) -- (x1t2);
\draw [dashed,thick] (x2) -- (y2);
\draw [enddirected]  (x2t4)--(1.5,1);
\draw [enddirected]  (y2)--(3.5,0.);
\draw [thick] (1.35,.35) -- (1.35,.65);
\draw [thick] (1.65,.35) -- (1.65,.65);
\end{tikzpicture}}}}
=0.
\end{eqnarray*}
\caption{The three 1-loop corrections to $\Delta(w)$ (without combinatorial factors). The dashed line is $\Delta(w)$, the bars are the spatial derivatives of the KPZ term; notations as in \cite{LeDoussalWieseChauve2003}.
 The first one $\delta_1 \Delta(w)$ contains the qEW terms. The second contribution $\delta_2 \Delta(w)\sim \lambda^2 \Delta(w)^2$ is new. The next two terms  $\delta_3 \Delta(w)\sim \lambda \Delta(w) \Delta'(w)$ cancel each other; they also vanish separately  since they are odd in $w$, whereas $\Delta(w)$ is even.}
\label{all-corrections-4-Delta}
\end{figure}

\subsection{The renormalized correlator $\Delta(w)$}

\label{s:The renormalized correlator Delta(w)}
In \Eq{Delta-def} we had defined the renormalized  (effective)  force correlator as 
\be
\label{Delta-eff}
\Delta(w-w'):=  m^4 L^d \overline{(u_w - w)(u_{w'}-w')}^c.
\ee  
The definition of $u_w$ is given in \Eq{eq:uw}.
This is the same definition as the one used for qEW
\cite{LeDoussalWiese2006a,MiddletonLeDoussalWiese2006,RossoLeDoussalWiese2006a,BonachelaAlavaMunoz2008}.
Integrating the equation of motion \eq{eq:qKPZ} over space for a configuration $u_w(x):=u(x,t)$ at rest yields 
\be\label{18}
m^2(w- u_w) + \frac {1}{L^d}\int_x  \underbrace{\lambda \left[\nabla u_w(x)  \right]^2 + F\big(x,u_w(x)\big)}_{\mbox{total force}}=0.
\ee
Thus the correlator in \Eq{Delta-eff} measures fluctuations of the {\em total force}. Only for qEW ($\lambda=0$) this equals the  force exerted by the disorder. 
To be specific, let us define
\bea
F_w &:=& \frac1{L^d} \int_x   F\big(x,u_w(x)\big), \\
\Lambda_w &:=& \frac 1{L^d}\int_x \lambda [\nabla u_w(x)]^2.
\eea
A configuration at rest then has 
\be
m^2 (w-u_w)+ F_w+ \Lambda_w = 0.
\ee
Our goal is to compare observables with objects  in the field theory. What is calculated there is the effective action, or more precisely its 2-time contribution. 
(In the statics this would be the 2-replica term.) It is the sum of all connected 2-time diagrams, i.e.\ with two external $\tilde u$ fields. 
To 1-loop order, these are shown in Fig.~\ref{all-corrections-4-Delta}. 
The 2-point function $\overline{ uu}^c$ is obtained to all orders by contracting the 2-time contribution to the effective action with two response functions. 
While in real space this is a convolution, in momentum and frequency space this is simply a multiplication with the response function $R(k,\omega)$.
According to \Eq{Delta-eff} it is  to  be evaluated at   momentum $k=0$ and frequency $\omega=0$.
Recall that the response function $R(x, t)$  is the response of the observable $u (x, t)$ to a small uniform kick in force $f$  at $(x,t)=(0,0)$. 
Since the center of mass follows the center of the driving parabola $w$, 
\be
f_{\rm c} =\overline {w-u_w}=\mbox{const}  \quad
\Rightarrow \quad \partial_w \overline {u_w} = 1.
\ee
Thus a uniform kick $f= m^2 \delta w$ leads to a response  for the center of mass according to $u_w \rightarrow u_w + \delta w =u_w + \frac{f}{m^2} $.
As a result, the integrated response function is given by
\be\label{intresp}
\int_t R(k=0,t) \equiv  \frac1{L^d}\int_x \int_t R(x,t) = \frac1{m^2}.
\ee
This is equivalent to $ R(k=0,\omega=0) =  m^{-2}$. 

We finally need to remember the field-theoretic definition of the effective action $\Gamma$: It is obtained from the corresponding expectation values by {\em amputation} of the response function, which is equivalent to dividing by the response function (in Fourier representation). Due to \Eq{intresp} this is nothing but multiplication with $m^2$, once for each of the two external fields $u$. This gives the factor of $m^4$ in \Eq{Delta-eff}, and \Eq{Delta-eff} is nothing but the 2-time contribution to the effective action $\Gamma$, equivalent to the renormalized   force  correlator $\Delta(w)$. 
It is the $(k=0, \omega = 0)$ mode of the full  effective force correlator  in the field theory for depinning.

Having established that \Eq{Delta-eff} is the proper definition of the renormalized $\Delta(w)$, 
it is still instructive to study the correlations of all three     forces appearing in   \Eq{18}.
To this aim, let us define in addition to \Eq{Delta-eff}
\bea\label{24}
\Delta_{FF}(w-w') &:=& L^d \overline{ F_w F_{w'}}^{\rm c}, \\
\Delta_{F\Lambda}(w-w') &:=& L^d \overline{ F_w \Lambda_{w'}}^{\rm c}, \\
\Delta_{\Lambda \Lambda}(w-w') &:=& L^d \overline{ \Lambda_w \Lambda_{w'}}^{\rm c}.
\label{26}
\eea
A measurement of these quantities is shown below in Fig.~\ref{fig:corqKPZ}.

Let us finally give the scaling dimensions, 
\be\label{dimDelta}
\Delta(0) \sim m^{4} \xi_{m}^{d} \left[u_{w} -w\right]^{2}  \sim m^{4-d \frac{\zeta_{m}}{\zeta}-2 \zeta_{m}}.
\ee
The scaling of the argument of $\Delta(w)$ is given by
\be
w \simeq u \sim  m^{-\zeta_{m}}.
\ee
These scalings are reflected in the FRG flow equations derived below in \Eq{eq:flowdelta}.

\subsection{Link to directed percolation,    exponents given in the literature, and other    relations}
\label{Link to directed percolation,    exponents given in the literature, and other    relations}

For TL92 in $d=1$, the scaling of a blocked interface at depinning is  given by directed percolation \cite{TangLeschhorn1992,AmaralBarabasiBuldyrevHarringtonHavlinSadr-LahijanyStanley1995,BarabasiGrinsteinMunoz1996,Hinrichsen2000,AraujoGrassbergerKahngSchrenkZiff2014,Dhar2017}.
In table \ref{d=1-num-table} we summarize the exponents obtained this way, which  guide us in  the construction and   tests of the FRG. Details are given in \cite{MukerjeeBonachelaMunozWiese2022}.

\begin{table}[t]
$\displaystyle\normalsize
\begin{array}{rclrcl}
\nu_{\parallel} &=& 1.733 847 (6),\qquad  & \qquad
\nu_{\perp} &=& 1.096 854 (4), \\
 \zeta &=& 0.632613 (3), &
\zeta_m &=&  1.046190  (4), \\
\frac{\zeta_m}{\zeta} &=& 1.65376(1),  & 
\tau &=& 1.259246(3), \\
\beta_{\rm dep} &=& 0.636993(7), &
\psi_\lambda &=& 0.26133(2),\\
\psi_k &=& 1.30752(2), & 
\end{array}
$
\caption{Numerical values for all exponents used in this section ($d=1$), as obtained from Ref.~\cite{Hinrichsen2000} combined with the  scaling relations derived here.}\label{d=1-num-table}
\end{table}

In dimensions $d\geq 2$  directed percolation paths are
1-dimensional, whereas the interface is $d$-dimensional. As a result,  the mapping to DP no longer exists,  and one has to introduce directed surfaces
\cite{BarabasiGrinsteinMunoz1996}.
 The exponents we find in  $d=2$ and $d=3$ are summarized on table \ref{tab:flow} (page \pageref{tab:flow}).

\subsection{The effective action in simulations}\label{sec:simul}
To guide our field-theoretical work, we first checked in dimension $d=1$ that the scaling exponents given in table \ref{d=1-num-table}   account for the measured values of $\psi_{\rm c}$ and $\psi_\lambda$ given in \Eqs{Psi-c}-\eq{Psi-lambda}.
To this aim,  a  novel algorithm was designed  \cite{MukerjeeBonachelaMunozWiese2022} to measure $\psi_{\rm c}$ and $\psi_\lambda$ by imposing a spatial modulation in the background-field configuration $w$. 
The simulations were performed for three different models, all in the qKPZ
universality class: the cellular automaton  TL92 \cite{TangLeschhorn1992}, anharmonic depinning \cite{RossoKrauth2001b,MukerjeeBonachelaMunozWiese2022},  and a direct simulation of \Eq{eq:qKPZ}  \cite{MukerjeeBonachelaMunozWiese2022}. The best results were achieved for 
  anharmonic depinning, thanks to an efficient algorithm for its  evolution \cite{RossoKrauth2001b}.

\begin{figure}
\mbox{\!\!\!\!\includegraphics[width=1.04\linewidth]{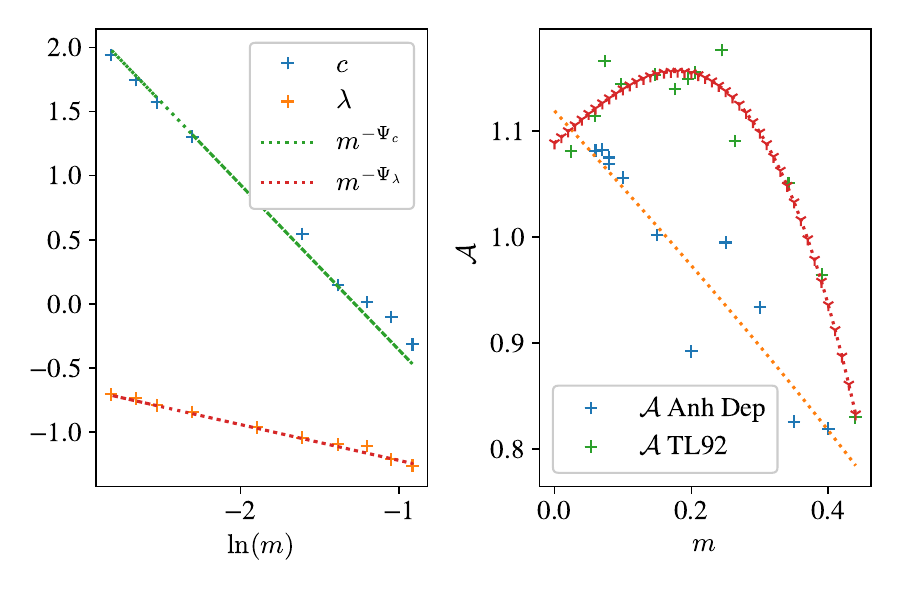}}
\caption{Left: Effective $c$ and $\lambda$  for anharmonic depinning. Right: Convergence to the fixed point as $m\rightarrow 0$,  both for anharmonic depinning and TL92.  The dotted lines are guides for the eye.}
\label{fig:fixedpoint}
\end{figure}

With the  novel algorithm designed in \cite{MukerjeeBonachelaMunozWiese2022}, we measured the effective couplings  $\lambda$ and $c$, as a function of $m$. In Fig.~\ref{fig:fixedpoint} (left) we show  their flow
as a function of $m$. 
To be specific, what we measure (left), and what is predicted from DP via table \ref{d=1-num-table} (right) is

\bea
\psi_{\rm c}^{ d=1} &=& 1.31(4), \qquad \psi_{\rm c}^{\rm DP} = 1.30752(2), \\
\psi_\lambda^{d=1} &=& 0.28(3), \qquad \psi_\lambda^{\rm DP} = 0.26133(2).
\eea
This confirms our scaling analysis and allows us to
measure as shown on  Fig.~\ref{fig:fixedpoint}   the dimensionless amplitude 
\be
 \mathcal{A}:= \frac{\Delta(0)}{| \Delta^{\prime}(0^+) |} \frac{\lambda}{c} .
 \label{eq:fixedpointA}
 \ee 
The ideas behind this definition is that the KPZ term has one field more than the elastic term. Thus the ratio $\lambda/c$ has the inverse   dimension of a field, which is compensated by the first ratio. 
That $\ca A$ converges to the same value for two different models gives strong evidence that qKPZ is the effective theory, and that   a fixed point of the renormalization-group flow is reached.
In $d=1$, this ratio reads
\be
\ca A^{d=1} = 1.10(2).
\ee
The last points to verify is that we can measure the effective-force correlator $\Delta(w)$, that different models in the qKPZ class have the same   $\Delta(w)$, and that this function is close to, but distinct from the one for qEW. This is shown in Fig.~\ref{fig:corsfrg1d}.

\section{Field theory}\label{sec:fieldtheory}

Now that we  verified that all models have a fixed point represented by the qKPZ equation,  and that we have the correct scaling dimensions for every variable, we can confidently  
  construct their field theory.

\begin{figure}
\centerline{\diagram{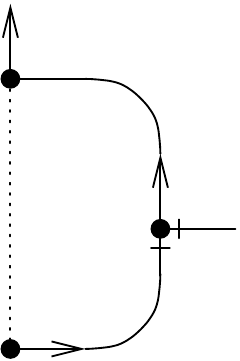}\qquad\diagram{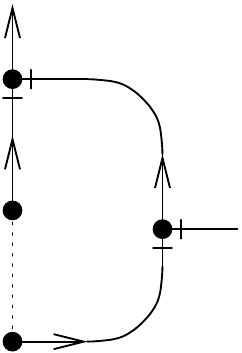}}
\caption{The 1-loop corrections to $c$.}
\label{fig:diags-c}
\end{figure}
\begin{figure}[t]
\centerline{\diagram{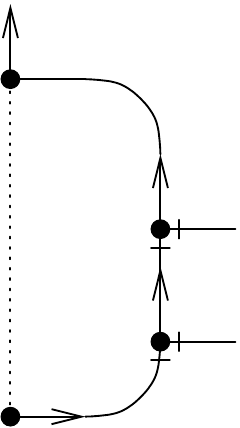}\qquad\diagram{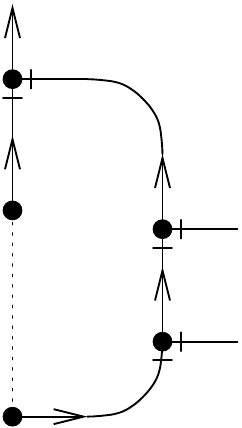}\qquad\diagram{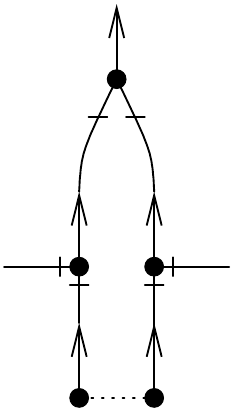}}
\caption{{1-loop diagrams correcting $\lambda $.}}
\label{fig:diags-lambda}
\end{figure} 

\subsection{Reminder: Generation of KPZ term from anharmonic elasticity}
\label{s:Generation of KPZ term from anharmonic elasticity}

Let us remind how anharmonic elastic terms generate a KPZ term at depinning \PK:
To this purpose consider a standard elastic energy, supplemented by an 
additional anharmonic (quartic) term (setting $c=1$ for simplicity), 
\begin{equation}\label{Eanharm}
\ca H_{\mathrm{el}}[u]= \int_{x} \frac{1}{2}\left[\nabla
u (x)\right]^{2}+\frac{c_{4}}{4} \left[\left(\nabla u (x)\right)^2\right]^{2}. 
\end{equation}
The corresponding terms in the  equation of motion read
\bea\label{lf28m}
 \partial_t u (x,t) &=&  \nabla^2 u (x,t) +c_4 \nabla \left\{\nabla u(x,t)\left[ \nabla u
(x,t)\right]^2 \right\}\nn\\
&& + ...
\eea
Since the r.h.s.\ of \Eq{lf28m} is a total derivative, it is surprising 
that a KPZ-term can be generated  in the limit of a {\em vanishing}
driving velocity. This puzzle was solved in Ref.~\cite{LeDoussalWiese2002}, where the KPZ term  arises by contracting the non-linearity with one  bare disorder (we drop the index on $\Delta_0$ from now on for simplicity of notation), 
\begin{eqnarray}
\!\!\!\delta \lambda &=& \parbox{2cm}{\pfig{2cm}{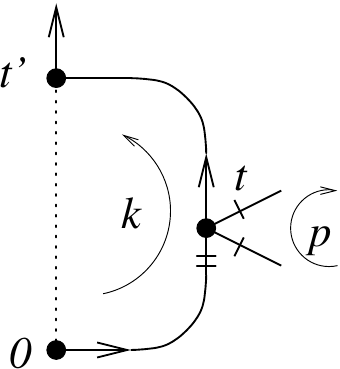}} \nn \\
&=& -\frac{c_{4}}{p^{2}} 
\int_{t>0} \int_{{t'>0}} \int_{k} \rme^{-(t+t')(k^2+m^2)} \left[k^{2}p^{2}+2
(kp)^{2}\right] \nonumber \\
&&\qquad \qquad \qquad \quad 
\times   \Delta'\big(u({x,t+t'})-u({x,0})\big) .\qquad
\label{k1}
\end{eqnarray}
As $u(x,t+t')- u(x,0)\ge 0$, the leading term in  \Eq{k1} can be written as 
\bea\label{lf2a}
\delta\lambda = - \frac{c_{4}}{p^{2}} \int\limits_{t>0}\int\limits_{t'>0}\int\limits_{k}
\rme^{-(t+t')(k^2+m^2)} 
 [k^{2}p^{2}{+}2 
(kp)^{2} ]  \Delta' (0^{+}). \nn\\
\eea
Integrating over $t,t'$ and
using the radial symmetry in $k$ yields
\begin{equation}\label{lf4}
\delta \lambda = - c_{4} \left(1+\frac{2}{d}
\right)\int_{k}\frac{\Delta' (0^{+}) k^2}{(k^{2}+m^2)^2}  . 
\end{equation}
This shows that in the FRG a KPZ term is generated from the non-linearity. As $-\Delta'(0^+)>0$, its amplitude is positive. The integral 
\eq{lf4} has a strong UV divergence, thus the generation of this term happens at small scales, similar to the generation of the critical force, see appendix \ref{a:Fc}.

\subsection{1-loop contributions} \label{s:1-loop contributions}
\begin{figure}[t]
\centerline{\diagram{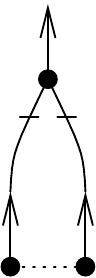}}\caption{{Additional 1-loop correction to  $\eta $ for qKPZ as compared to qEW.}}
\label{fig:remaining-diagrams}
\end{figure}

Here we summarize the 1-loop contributions to $c$, $\lambda$, $\eta$ and $\Delta$. This is {\em almost}  the same calculation as in Ref.~\cite{LeDoussalWiese2002}, with a little twist: Since we   work in a massive scheme, many of the cancelations in 
\cite{LeDoussalWiese2002} no longer exist. We remind that this change in scheme was forced upon us by our decision to measure the effective parameters of the theory, necessitating to drive with a confining potential.  We believe that this is also much closer to real experiments. 
It is a scheme widely used for perturbative RG for the Ising model in $d=3$, pioneered by G.~Parisi and used up to 7 loop-order by B.~Nickel and collaborators \cite{ParisiBook,Parisi1980,BakerNickelGreenMeiron1976,NickelMeironBaker1977,BakerNickelMeiron1978}.
As discussed above, we think of  this fixed-dimension renormalization scheme as an expansion around the $d=0$ qEW fixed point.
The diagrams   from the perturbation in $\lambda$ are given   in Figs.~\ref{fig:diags-c}-\ref{fig:remaining-diagrams}.

We obtain the same diagrams as in \cite{LeDoussalWiese2002}  but with   coefficients $a_i $  that differ from \cite{LeDoussalWiese2002}   away from the upper critical dimension. 
The explicit calculations are  given in appendix \ref{sec:diagrams}. Terms with numerical coefficients only (no $a_i$) are those appearing already in qEW. 
\bea
\label{delta-eta/eta}
\frac{\delta \eta}{\eta} &=&-\left[ a_0   \hat \lambda \Delta^{\prime}\left(0^{+}\right)+ \Delta^{\prime \prime}\left(0^{+}\right)\right] I_1, \\
\label{delta-c/c}
\frac{\delta c}{c}&=&-\left[a_1  \hat \lambda   \Delta^{\prime}\left(0^{+}\right)+a_2 \hat \lambda^{2}   \Delta(0)\right] I_1, \\
\frac{\delta \lambda}{\lambda}&=&-\left[a_3 \hat \lambda   \Delta^{\prime}\left(0^{+}\right)+a_4 \hat \lambda^{2}   \Delta(0)\right] I_1, 
\label{delta-lambda/lambda}\\
\delta \Delta(u)&=&\left\{  a_{5} \hat \lambda^{2}   \Delta(u)^{2}- \partial_u^2 \frac 12 \left[ \Delta(u)-\Delta(0) \right]^2 \right\} I_1,\qquad
\label{delta-Delta/Delta}
\eea
\bea
\!\!\! \hat \lambda &:=& \frac \lambda c, \\
\label{I1def}
\!\!\!I_1 & =& \int_k \frac{1}{(c k^2 + m^2)^2}, \\
\!\!\!a_0 &=& \frac{d}{4}, \qquad  a_1 = 1, \qquad  a_2 = \frac{d-1}{3}, \\
\!\!\! a_3  &=& 1, \qquad a_4= \frac{d+2}{6},\qquad  a_5 = \frac{d(d+2)}{12} .\qquad   
\eea
The coefficients $a_i$ in the limit of $d\to 4$ used by \cite{LeDoussalWiese2002} are obtained by setting $d\to 4$, resulting into $a_i=1$ for all $i$,  except   $a_5 = 2$. While this is the standard procedure followed in a dimensional expansion,  it misses that in dimension $d=0$ the KPZ term does not exist, thus cannot correct the remaining terms: viscosity $\eta$, and  effective force correlator $\Delta(u)$. The factors of $d$ in coefficients $a_0$ and $a_5$ reflect this  physical necessity. No such constraint exists for $c$ and $\lambda$: since they are absent from the equation of motion \eq{eq:qKPZ} in $d=0$, their coefficients can well be modified. 

As $\lambda$ and $c$   appear   in the combination of $\hat \lambda = \lambda/c$, the important question is whether this ratio is corrected. This is indeed the case as  
\be\label{37}
\frac{\delta \hat \lambda}{\hat \lambda} = (a_2-a_4) \hat \lambda^2 \Delta(0) I_1 = -\frac{4-d}6 \hat \lambda^2 \Delta(0) I_1 .
\ee
Note that this term is negative, and have a power  in $\hat \lambda$ superior to one. It will therefore stop the RG flow for $\hat \lambda$ at large $\hat \lambda$, allowing us to close our system of equations!

A final important point to mention is that the confining potential $\sim m^2$ is not renormalized. In qEW this is due to the statistical tilt symmetry (STS) \cite{Wiese2021}, which can be checked perturbatively: Since the  effective force correlator  contains $u$ only as a difference $u(x,t)-u(x,t')$, no field $u$ without a time derivative can be generated. The same holds true  here: since the additional KPZ vertex has additional spatial derivatives, it cannot generate a field $u$ without spatial derivatives. This property is very useful, as we can as in qEW use $m$ as an RG scale, without  caveat.

Finally, the critical force  is
\bea\label{Fc-main-text}
F_{\rm c} &= &F_{\rm c}^{(1)}+F_{\rm c}^{(2)} \nn\\
&\simeq&  \left[  \Delta'(0^+) + \frac d 2   \hat \lambda   \Delta(0)\right] \int_k \frac1{c k^{2}   +m^2}. 
\eea
The first contribution is negative, identical to qEW. The second is positive, and specific to qKPZ.  The  non-linearity reduces the force needed to depin the interface.
This is derived in appendix \ref{a:Fc}.

\subsection{Flow equations}
\label{s:Flow equations}
Above we calculated  the perturbative  corrections. We now derive the corresponding RG relations. 
Since $m$ is
not corrected under renormalization, we   use it to parameterize the flow of the remaining quantities. 
To this aim, first define the dimensionless field  as
\be
\mathbf{u}:= u \,m^{\zeta_m}. 
\ee
We have $-m \frac{\partial}{\partial m}\Delta(u)  = [\delta \Delta] \varepsilon I_1 $.
The integral $I_1$ defined in \Eq{I1def}
is evaluated in  \Eq{a:I1} of appendix \ref{a:Useful momentum integrals}, 
\bea\label{I1-main-text}
I_1  &:=& \int_k \frac{1}{(c k^2 + m^2)^2} = \frac{m^{d-4}}{c^{d/2}} \frac{2\Gamma(1+\frac \epsilon 2)}{\epsilon (4\pi)^{d/2}}.~~~
\eea 
It scales as 
\bea
&I_1 \sim \xi_m^{-d} m^{-4}, 
\eea
where we remind that
\be
\Delta(0) \sim \xi_m^d m^4 u^2.
\ee 
The {\em dimensionless  renormalized correlator} $\tilde{\Delta}(\mathbf{u})$ is then defined in terms of the effective  force  correlator $\Delta(u)$, such that it absorbs $\varepsilon I_1$ as 
\be\label{Delta-reno}
\tilde{\Delta}(\mathbf{u}):= \epsilon I_{1}  m^{2\zeta_m} \Delta\left(u=\mathbf{u} m^{-\zeta_m}\right). 
\ee
The explicit $m$-dependent factor in front of $\Delta$ is the scaling dimension given in \Eq{dimDelta}. 
This yields the flow equation for the effective dimensionless   force correlator,
\be
\begin{aligned}
\partial_{\ell} \tilde{\Delta}(u)=&\left(4-d \frac{\zeta_{m}}{\zeta}-2 \zeta_{m}\right) \tilde{\Delta}(u)+u \zeta_{m} \tilde{\Delta}^{\prime}(u) \\
&+\frac{d(d+2)}{12} \tilde{\lambda}^{2} \tilde{\Delta}(u)^{2} \\
&-\tilde{\Delta}^{\prime}(u)^{2}-\tilde{\Delta}^{\prime \prime}(u)\big[ \tilde{\Delta}(u)-\tilde{\Delta}(0)\big] .
\end{aligned}
\label{eq:flowdelta}
\ee
Here we defined  the dimensionless combination $\tilde \lambda$
\be
\tilde \lambda:= \frac{\lambda}{c} m^{-\zeta_m}  \equiv \hat \lambda  m^{-\zeta_m}.
\ee
Its flow equation is obtained from \Eq{37} as 
\be
-   {m} \partial _m   {\tilde \lambda} = \zeta_m \tilde \lambda - \frac{4-d}6  \tilde \lambda ^3 \tilde \Delta(0) .
\ee
It has one fixed point $\tilde \lambda=0$, and a second  non-trivial fixed point at 
\be\label{e:lambda-c}
\tilde \lambda_{\rm c}  = \sqrt{ \frac{6 \zeta_m}{(4-d)\tilde \Delta(0)}}. 
\ee
We can see that in $d=4$ the fixed point disappears as $\tilde{\lambda} $ goes to infinity.

\noindent
The anomalous dimension $\psi_{\rm c}$ defined in \Eq{Psi-c} reads 
\be
\psi_{\rm c} =   -\tilde \lambda \tilde \Delta'(0^+) -\frac{d-1}{3} \tilde \lambda ^2  \tilde \Delta(0) . 
\ee
Using \Eq{9}, we find
\be
\frac{\zeta_{m}}{\zeta}  =  1 +\frac{1}{2} \left[-\tilde{\lambda} \tilde{\Delta}^{\prime}\left(0^{+}\right)-\frac{d-1}{3} \tilde{\lambda}^{2} \tilde{\Delta}(0)\right].
\label{eq:fixedpointc}
\ee
\Eq{eq:flowdelta} is still cumbersome to solve. Reinjecting  \Eq{eq:fixedpointc}, we  obtain at the fixed point
\bea
 0 &=&\left(\epsilon +\frac{d}{2}\left[\tilde{\lambda} \tilde{\Delta}^{\prime}\left(0^{+}\right)+\frac{d-1}{3} \tilde{\lambda}^{2} \tilde{\Delta}(0)\right]-2 \zeta_{m}\right) \tilde{\Delta}(u) \nn \\
&&+u \zeta_{m} \tilde{\Delta}^{\prime}(u)+\frac{d(d+2)}{12} \tilde{\lambda}^{2} \tilde{\Delta}(u)^{2} \nn \\
&&-\tilde{\Delta}^{\prime}(u)^{2}-\tilde{\Delta}^{\prime \prime}(u)\big[\tilde{\Delta}(u)-\tilde{\Delta}(0)\big]. 
\label{eq:fixedpointdelta}
\eea
The anomalous contribution $\psi_\eta$ reads \be
\psi_\eta =  -\left[ \frac{d}{4} \tilde{\lambda} \tilde{\Delta}^{\prime}\left(0^{+}\right)+ \tilde{\Delta}^{\prime \prime}\left(0^{+}\right)\right].
\ee
Using \Eq{11} this yields
\bea
\label{eq:fixedpointeta}
z&=& \frac{\zeta}{\zeta_m}\left[ 2 - \frac{d}{4} \tilde{\lambda} \tilde{\Delta}^{\prime}\left(0^{+}\right)- \tilde{\Delta}^{\prime \prime}\left(0^{+}\right)\right].
\eea 
We note that for $d \rightarrow 0$ the contribution of $\tilde \lambda$
in equation \eq{eq:fixedpointdelta} disappears, thus we recover the qEW fixed point. This is not the case in the massless  scheme  \PK. 
Increasing $d$ we expect the qKPZ fixed point to smoothly move away from the qEW one. 
In Figure  \ref{fig:corsfrg1d} we show that in dimension $d=1$ the shape of the {\em measured}  $\Delta(w)$ for qEW and qKPZ are close, even though their amplitudes may be rather different. We take this as  an encouraging sign to construct the FRG fixed point for qKPZ. This is the task of  section  \ref{solution-4-flow}. Since our expansion is uncontrolled, we need to obtain additional safeguards in order to see if where our approach hold, and where it is too crude. For that, we derive constraints to be satisfied by the fixed point.

\subsection{Necessary conditions for a fixed point, and bounds}
\label{s:Necessary conditions for a fixed point, and bounds}
\subsubsection{Disorder and   force correlator  relevant}
We now assume   (as in qEW) that the  effective force correlator is relevant, thus $4-d\frac{\zeta_m}{\zeta}- 2 \zeta_m >0$. This is satisfied in $d=1$,   see Table \ref{d=1-num-table}.  There one finds $4-d\frac{\zeta_m}{\zeta}- 2 \zeta_m=0.253859$.
To compare,  in $d=0$ (qEW) one gets $4-2\zeta_m = 4-2\times 2^-\approx 0$. In $d=1$ qEW has $4-1-2\times 5/4=0.5$.

Taking the limit of $u\to 0$ in \Eq{eq:fixedpointdelta}, we obtain a 
  soft bound at 1-loop order, 
\be\label{eq:nec-cond-FP2}
|\tilde \Delta'(0^+)| > \sqrt {\frac{d(d+2)}{12}} \tilde \lambda \tilde \Delta(0).
\ee
When violated,  the rescaling term becomes negative, and we expect the  effective force correlator to disappear at large scales. 
Using the definition of the universal amplitude  $\ca A $ in \Eq{eq:fixedpointA}, we can rewrite the bound \eq{eq:nec-cond-FP2} as\footnote{Note that the definition \eq{eq:fixedpointA} for $\ca A$ remains unchanged upon replacing all quantities by their dimensionless analogue, noted with a tilde.}
\be \label{fc-bound}
 \mathcal A < \ca A_{\rm c}^\Delta =\sqrt{\frac{12}{d(d+2)}} = \left\{\begin{array}{c} 
2 \mbox{~in~} d=1\\
1.22 \mbox{~in~} d=2\\
0.894\mbox{~in~} d=3
\end{array}
\right. ~.
\ee
\subsubsection{$\zeta_m>\zeta$}
We expect that the effective $c$ would grow at large scales, since it describes the long distance behavior of models with stronger than harmonic elasticity.
As a result we demande that  $\psi_{c}>0$ (which implies $\zeta_m>\zeta$ ).  Eq.~\eq{eq:fixedpointc} then yields 
\be
\tilde \lambda\times \left[    \tilde \Delta'(0^+) + \frac{d-1} 3 \tilde \lambda  \tilde \Delta(0) \right] < 0.
\ee
This can be rewritten as 
\be
\ca A<  \ca {A}_{\rm c}^{\psi_{c}}= \frac 3{d-1} .
\ee

\begin{figure}[t]
\centering
\includegraphics[width=\linewidth]{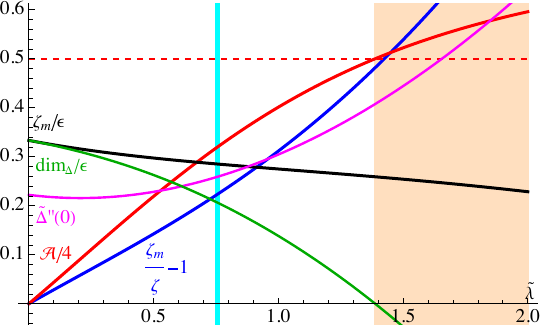}
\caption{In $d=1$: The 1-loop contributions $\zeta_m/\epsilon$, amplitude ratio $\ca A$ and $\zeta_m/\zeta-1$ as a function of $\tilde \lambda$. Setting $d=1$ in the flow equations. The orange shaded range is excluded by demanding that $\Delta$ is relevant, the cyan line is the location of the fixed point for $\tilde \lambda$. The red dashed line is the bound on $\ca A$ from $\ca A_{\rm c}^{\Delta} = \ca A_{\rm c}^{f_{\rm c}}$. (see section \ref{s:Positive pinning force})}
\label{fig:phasediag1}
\end{figure}

\subsubsection{Positive pinning force}
\label{s:Positive pinning force}
The last  condition is that the critical  force at depinning needs to be negative (keeping  us pinned), equivalent to a negative square bracket in \Eq{Fc-main-text}. In terms of $\ca A$, this results in
\be\label{bound-fc}
 \ca A \le \ca A_{\rm c}^{f_{\rm c}} =\frac 2 d.
\ee
We find that in $1\le d\le 4$ the strongest bound is $\ca A_{\rm c}^{f_{\rm c}}$ for the critical force, followed by the one for $\Delta(w)$ and  $\psi_{c}$, 
\be\label{bounds}
\ca A < \ca A_{\rm c}^{f_{\rm c}} \le \ca A_{\rm c}^{\Delta} < \ca A_{\rm c}^{\psi_{c}} . 
\ee
It would be interesting to continue this to 2-loop order.

\subsection{Solution of the flow equations}

\label{solution-4-flow}

\subsubsection{Scheme}
\label{Scheme}

How do we solve these coupled equations (Eqs.~\eq{e:lambda-c}-\eq{eq:fixedpointeta} )
The procedure is adapted from the standard ansatz for qEW \cite{LeDoussalWieseChauve2002}, explained in detail in Ref.~\cite{Wiese2021}: 
\begin{enumerate}
\item[(i)] Use the normalization
$\tilde{\Delta}(0) = \epsilon $. 
In practice, this corresponds to setting $\epsilon\to 1$ and $\zeta_m \to \zeta_m/\epsilon$ in \Eq{eq:fixedpointdelta}, and then solving the flow equations  with $\tilde{\Delta}(0) \to 1$ in the code. 

\item[(ii)] Solve the (such rescaled) flow equation 
 \eq{eq:fixedpointdelta}  for
 $0 \le \tilde{\lambda} \le 2$. The correct solution is the one for which $\tilde \Delta(w)$ decays to zero at least exponentially fast:
 A power-law decay, or an increase with $w$, is not permitted by the physical initial condition. 

\item[(iii)] The  critical  $\tilde \lambda_{\rm c}$ that satisfies \Eq{e:lambda-c} in our scheme is
\be\label{lambda-c}
\tilde{\lambda}_{c} = \sqrt{\frac{{6}}{4-d}} \sqrt{\frac{ \zeta_m}{\epsilon}}.
\ee
Given $d$, the first square root is a number; the second one is   the result from step (ii) above. 
\end{enumerate}
It is interesting to see how the different exponents depends on $\tilde \lambda$ that is why we solve the flow equations for different $\tilde \lambda$ instead of plugging the value given by \Eq{e:lambda-c}.
 
\subsubsection{$d=1$} 
\label{d=1}
The procedure and the values obtained for different $\tilde \lambda$ are shown for $d=1$ in Fig.~\ref{fig:phasediag1}.
We see that $\zeta_m/\epsilon$   slightly decreases from its qEW value of $\zeta_m^{\rm qEW}=1/3$. 
The ratio $\zeta_m/\zeta$ starts at $1$ for $\tilde \lambda =0 $, and then   grows. The  effective force correlator  becomes irrelevant for $\tilde \lambda \approx 1.4$. At the same time the bound \eq{fc-bound} for $\ca A$ (marked here as a red dashed line $\ca A/4=0.5$) is violated. 
The critical $\lambda_{\rm c} = 0.755203$ respects all   bounds in \Eq{bounds}.  It gives 
\bea\label{60}
\zeta_m^{d=1} &=& 0.8555, \\
\label{61}
\zeta^{d=1} &=& 0.6994, \\
z^{d=1} &= & 1.2736, \label{62}\\
\ca A^{d=1} &=& 1.2781. \label{A-d=1}
\eea
This can be compared to their values for $\lambda=0$ (qEW), $\zeta_m=\zeta=1$, and $z=4/3$, and the numerically obtained values 
$\zeta_m=1.052$, $\zeta=0.636$, and $z=1.1$. The values \eq{60}-\eq{62} are pretty reasonable for 1-loop estimates: For qEW $\zeta$ in $d=1$ comes out $20\%$  smaller (1 instead of $1.25$); the same reduction applies to our prediction for $\zeta_m$ in qKPZ. $\zeta$ is about $10\%$   larger than the numerical value. Finally, while $z$ is too large, using the numerically known value for $\zeta/\zeta_m$ with the same 1-loop estimate would yield $z=0.942$, smaller than the measured value of $z=1.1$.  (Note that the  prediction of $z=1$  in \cite{AmaralBarabasiBuldyrevHarringtonHavlinSadr-LahijanyStanley1995} is invalidated by numerics  \cite{MukerjeeBonachelaMunozWiese2022}.)

\begin{figure}
\centering
\includegraphics[width=\linewidth]{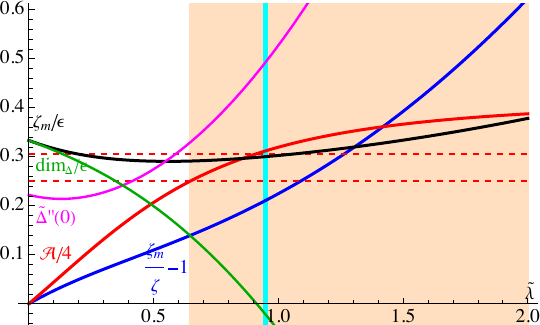}
\caption{Same as Fig.~\ref{fig:phasediag1} for $d=2$.  The lower red dashed line is the bound on $\ca A$ from $\ca A_{\rm c}^{f_{\rm c}}$, the upper one the bound from $\ca A_{\rm c}^{\Delta} $.}
\label{fig:phasediag2}
\end{figure}

\subsubsection{$d=2$}
\label{d=2}
Relevant quantities as a function of $\lambda$ are given on Fig.~\ref{fig:phasediag2}.
Evaluation at $\lambda=\lambda_{\rm c}$ yields
\bea\label{64}
\zeta_m^{d=2} &=& 0.6051, \\
\label{65}
\zeta^{d=2} &=& 0.4941, \\
z^{d=2} &= & 1.4112, \label{66}\\
\ca A^{d=2} &=& 1.2479.\label{67}
\eea
These results violate the bound \eq{bound-fc} on $\ca A$ for  $f_{\rm c}$. Supposing that this is an artifact of the 1-loop approximation, the next bound to consider is the bound \eq{fc-bound}, asking that  the  effective force correlator  is relevant at the transition. This bound is only slightly violated. 
We   therefore hope that the values given in \Eqs{64}-\eq{67} are   usable. 

Our own numerical simulations \cite{MukerjeeBonachelaMunozWiese2022} give $\zeta_m=0.70(3)$, $\zeta=0.47(3)$ for TL92, 
and  $\zeta_m=0.61(2)$, $\zeta=0.48(2)$  for anharmonic depinning. We expect the latter to be more reliable as there are less finite-size corrections. The agreement is then excellent. 

For comparison we   note that 1-loop qEW gives $\zeta_m=\zeta=2/3$, and $z=1.5556$, while numerics gives $\zeta=\zeta_m=0.753(2)$ and $z=1.56(6)$.

\subsubsection{$d=3$}
\label{d=3}
\begin{figure}
\centering
\includegraphics[width=\linewidth]{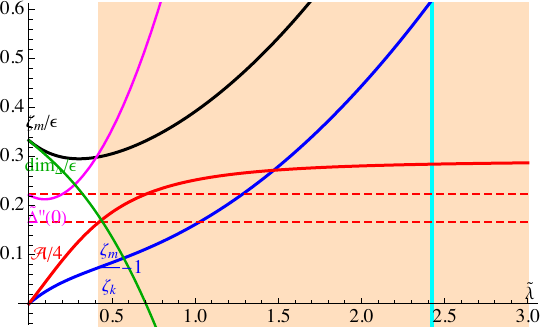}
\caption{Same as Fig.~\ref{fig:phasediag1} for $d=3$.  The lower red dashed line is the bound on $\ca A$ from $\ca A_{\rm c}^{f_{\rm c}}$, the upper one the bound from $\ca A_{\rm c}^{\Delta} $.}
\label{fig:phasediag3}
\end{figure}

Relevant quantities as a function of $\lambda$ are given on Fig.~\ref{fig:phasediag3}.
At the non-trivial fixed point \eq{lambda-c} for $\lambda$, we find
\bea\label{70-bis}
\zeta_m^{d=3} &\stackrel ? =& 0.9799, \\
\label{71}
\zeta^{d=3} &\stackrel ?=& 0.6048, \\
z^{d=3} &\stackrel ?= & 0.9777, \label{72}\\
\ca A^{d=3} &\stackrel ?=& 1.1394. \label{A-d=3}
\eea
These values violate all bounds, and thus need to be rejected. There are four possible conclusions:
\begin{itemize}
\item[(i)] since the  effective force correlator  is irrelevant at this fixed point, there is no qKPZ class.
\item[(ii)] this fixed point is irrelevant, but  there is a another fixed point not contained in our approach.
\item[(iii)] our approach is too crude.
\item[(iv)] our approach is crude as the fixed-point value for $\lambda$ is too large, but providing a better value for $\lambda_{\rm c}$ it   remains   predictive. 
\end{itemize}
If we believe Ref.~\cite{RossoHartmannKrauth2002},   there is a distinguished fixed point for both classes, eliminating (i) while allowing for (ii).  
While the following option (iii) is suggestive, we can still try (iv): we use $\lambda$ such that the   effective depinning force  at the fixed point is zero. Since the KPZ term grows under renormalization, it will finally render all pinned configurations unstable. This in turn reduces the generation of the KPZ term, making it   less relevant. Our conjecture, which needs to be validated in numerical simulations, is that the system   gets stuck at this precise point.  
Under this assumption we obtain
\bea\label{78}
\zeta_m^{d=3} &=& 0.2998, \\
\label{79}
\zeta^{d=3} &=& 0.2751, \\
z^{d=3} &= & 1.7620, \label{80}\\
\ca A^{d=3} &=&  0.6667. \label{A-d=3-new}
\eea
These values are pretty much in line with the simulations for anharmonic depinning in $d=3$:  $\zeta_m=0.34(3) $,  $\zeta=0.27(3)$. 
We do not know the values of $z$ and $\ca A$. 

  We   remark that the behavior in $d=3$ calls for more investigation: for example,  $d_{\rm c}^{\rm qKPZ}$ may be between $3$ and $4$.

\subsubsection{Force amplitude ratio}\label{sec:force_ratio}

Let us now address the relative fluctuations of forces defined in \Eqs{24} to \eq{26}.
At leading order in perturbation theory we can estimate from Fig.~\ref{all-corrections-4-Delta} (where the $\delta_i \Delta(w)$ are defined)  that 
\bea
 \frac{\Delta_{FF}(w) }{\Delta(w)} &\approx  &  \frac{\delta_1 \Delta(w) }{\delta_1 \Delta(w) +\delta_2 \Delta(w) +\delta_3 \Delta(w) },  \\
\frac{\Delta_{\Lambda\Lambda}(w) }{\Delta(w)} &\approx& \frac{\delta_2 \Delta(w) }{\delta_1 \Delta(w) +\delta_2 \Delta(w) +\delta_3 \Delta(w) },\\
 \frac{\Delta_{\Lambda F}(w) }{\Delta(w)} &\approx&   \frac{\delta_3 \Delta(w) }{\delta_1 \Delta(w) +\delta_2 \Delta(w) +\delta_3 \Delta(w) }   .\qquad 
\eea
These equations simplify upon using that $\delta_3 \Delta(w)=0$. 
Given the similar  functional  forms shown in Fig.~\ref{fig:corqKPZ}, let us focus on the relative amplitudes. 
With the universal amplitude $\ca A$ defined in \Eq{eq:fixedpointA},  we get
\bea
 \frac{\Delta_{FF}(0) }{\Delta(0)} &\approx& \frac{1}{1-\frac{d(d+2)}{12}\ca A^2},  \\
 \frac{\Delta_{\Lambda\Lambda}(0) }{\Delta(0)} &\approx& \frac{\frac{d(d+2)}{12}\ca A^2}{1-\frac{d(d+2)}{12}\ca A^2},\\
  \frac{\Delta_{\Lambda F}(0) }{\Delta(0)} &\approx& 0.
\eea\begin{figure}
\centering
\includegraphics[width=\linewidth]{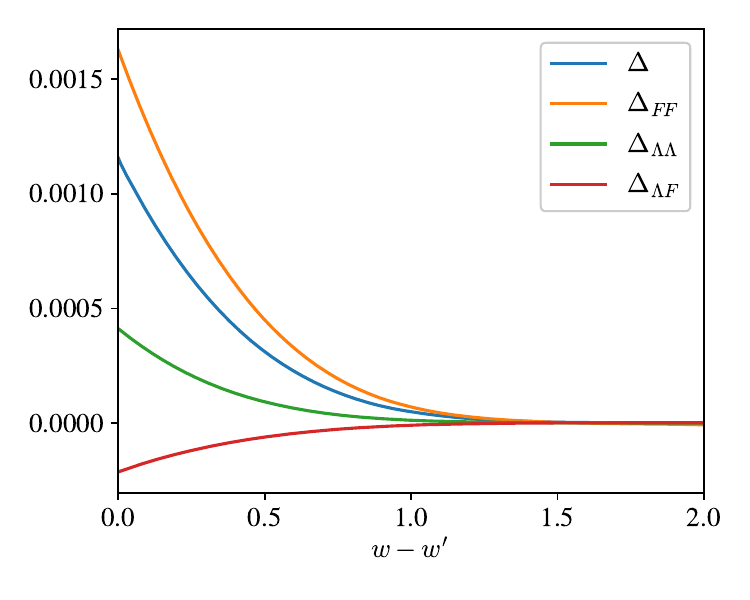}
\caption{Correlators of the disorder force, the interface center of mass, and the KPZ force, as well as the cross correlator of the KPZ force and the disorder force. The interface center of mass correlator is a mix of the disorder force and the KPZ force.}
\label{fig:corqKPZ}
\end{figure}In our simulations  in $d=1$ we find 
\bea \label{33}
 \frac{\Delta_{FF}(0) }{\Delta(0)}&=&1.40(3), \\
 \label{34}
 \frac{\Delta_{\Lambda\Lambda}(0) }{\Delta(0)} &=& 0.36(3), \\  \label{35}
  \frac{\Delta_{\Lambda F}(0) }{\Delta(0)} &=& -0.18(3).
\eea
The theory  in $d=1$ has
\be
\frac{d(d+2)}{12}\ca A^2 = 0.408,
\ee
which gives $1.69$, $0.24$ and $0$ for the three ratios in \Eqs{33} to \eq{35}.
Using the measured amplitude $\ca A=1.1$ these ratios become
$1.43$, $0.21$ and $0$ which is closer to the measured amplitudes. All these values seem pretty reasonable given the order of approximation.

\subsubsection{Other quantities and summary}
\label{Other properties}

Other properties of $\tilde \Delta (w)$  derived from the FRG solution are presented in table \ref{tab:curvature}. 
An interesting property is the curvature $\kappa$, defined as   
\bea
f(w)&:=& \ln \big(\Delta(w)/\Delta(0)\big), \nn\\
\kappa &:=& \frac 12 \frac{f''(0^+)}{f'(0^+)^2} = \frac12 \left[1 - \frac{\Delta(0) \Delta''(0^+)}{\Delta'(0^+)^2} \right] .
\eea 
It is constructed such that an exponential decaying $\Delta(w)$, which gives a straight line for $f(w)$, has a vanishing curvature.  
The definition was motivated by the observation in \PK\ that the FRG flow in the massless scheme possesses an exponentially decaying subspace, protected to all orders in perturbation theory. Our simulations in \us\ showed no evidence for this subspace. 
Still, $\kappa$ is a scale-free parameter which allows one to distinguish different shapes.

\begin{table}[b]\begin{tabular}{l@{~~}|@{~~}c@{~~}|@{~~}c@{~~}|@{~~}c@{~~}|@{~~}c@{~~}|@{~~}c@{~~}}
\textrm{quantity} &
$d$ &
qKPZ FT & qKPZ sim & qEW FT 
\\
\hline
$\kappa$ & $1$ &  $0.1291$ & $0.12(1)$ & $ 0.1667$ \\
& $2$ & $0.0738$ & $0.07(1)$ &$ 0.1667$  \\
& $3$ & $ 0.07704^*$ & $0.08(3)$& $ 0.1667$   \\
\end{tabular}
 \caption{Correlator quantities coming from the analytical solution of the flow equations, setting $\tilde \Delta(0)=\epsilon$. For qKPZ  in $d=3$ we fix $\tilde \lambda$ by supposing that the  effective force correlator  is marginal; the resulting values are  indicated by an asterisk. }
 \label{tab:curvature}
\end{table}

\begin{figure*}
\centerline{\includegraphics[width=1\textwidth]{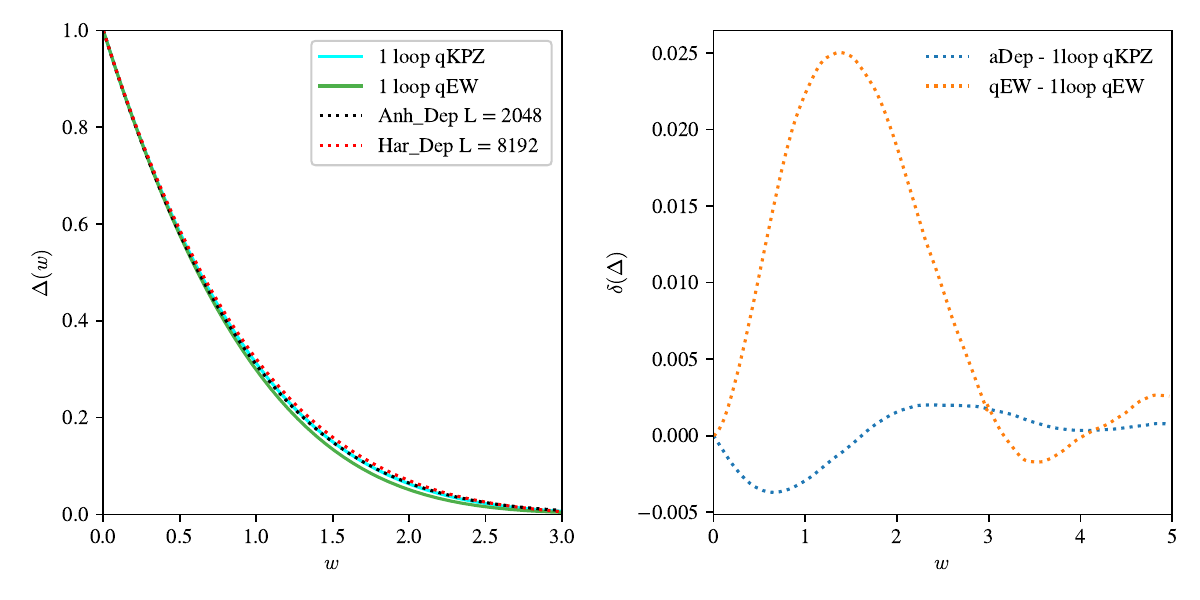}}
\caption{(colors online) (Left) Correlators in $d=1$ from simulations of harmonic depinning (qEW) and anharmonic depinning (in the qKPZ universality class), compared to the analytic solution of the flow equations. $\Delta(w)$ for anharmonic depinning decays slightly faster than the one for harmonic depinning.  The correlators are rescaled such that $\Delta(0)=|\Delta'(0^+)|=1$. (Right) Difference of the rescaled correlators measured or analytical. The qKPZ FRG 1-loop solution is around three times closer to the numerical simulation than the same curves for qEW.
}
\label{fig:corsfrg1d}
\end{figure*}

\begin{figure}
\includegraphics[width=1.0\linewidth]{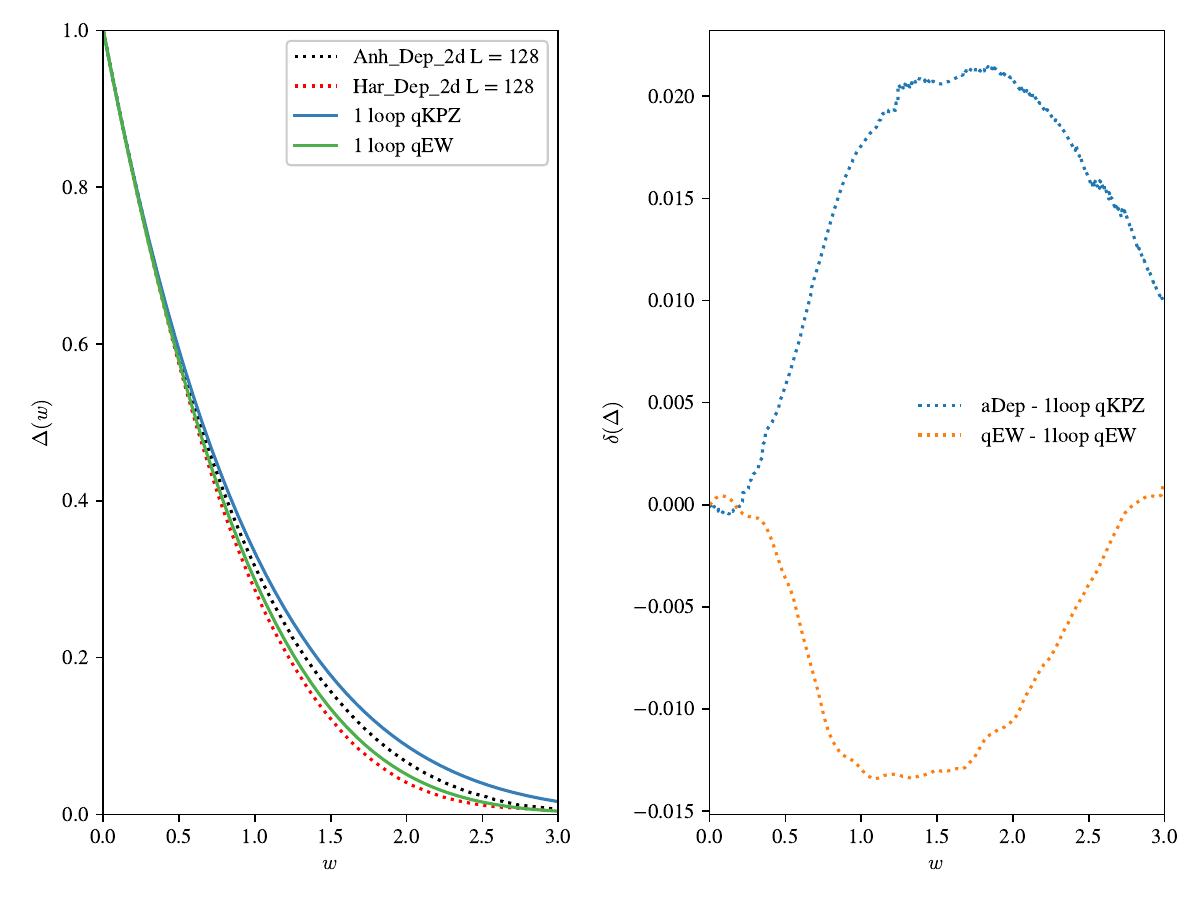}
\caption{(Left) Correlators in $d=2$ from simulations of harmonic depinning (qEW) and anharmonic depinning (qKPZ class), compared to the   solution of the FRG flow equations. The FRG solution is much closer to anharmonic depinning than to qEW. The correlators are rescaled such that $\Delta(0)=|\Delta'(0^+)|=1$. (Right)  Difference of the rescaled correlators measured and analytical. The agreement between simulations and theory is of the same order of magnitude for the two universality class, even if the qKPZ theory is much more sophisticated.}
\label{fig:corsfrg2d}
\end{figure}

Our results for the exponents are summarized in table~\ref{tab:flow}, and in  Figs.~\ref{fig:corsfrg1d} and \ref{fig:corsfrg2d}  for the full function $\tilde \Delta(w)$, rescaled such that $\tilde \Delta(0)=-\tilde \Delta'(0^+)=1$.  They show  excellent agreement between theory and simulation.

\begin{table}[h]\begin{ruledtabular}
\begin{tabular}{@{}llllll@{}}
\textrm{Exponent}&
\textrm{dim}&
\textrm{field theory}&
\textrm{simulations}\\

\colrule
$\zeta$ & $1$ & $0.6994$ & $0.636(8)$ \\
& $2$ & $0.4941$ & $0.48(2) $ \\
& $3$ & $0.2751^*$ & $0.27(3)$  \\
\hline
$\zeta_m$ & $1$ & $0.8555$ & $1.052(5) $ \\
& $2$ & $0.6051$ & $0.61(2) $ \\
& $3$ & $0.2998^* $& $0.34(3)$ \\
\hline
$z$ & $1$ & $1.2736$ & $1.10(2)$  \\
& $2$ & $1.4112$ & \\
& $3$ & $1.762^*$ & \\
\hline
$\mathcal{A}$ & 1 & 1.2781 & $ 1.1(1) $  \\
& $2$ & $1.2479$ & \\
& $3$ & $0.6667^*$ &
\end{tabular}
 \caption{Critical exponents of the qKPZ class, from simulations of anharmonic depinning   (except for $z$  coming from TL92) and the analytical resolution of
 the fixed-point equations. In $d=3$ we fix $\tilde \lambda$ by supposing that the  depinning force remains positive, indicated by an asterisk. Note that our simulations agree with  \cite{RossoHartmannKrauth2002}, and with the static exponents of \cite{AmaralBarabasiBuldyrevHarringtonHavlinSadr-LahijanyStanley1995} for $d\leq 2$, see \cite{MukerjeeBonachelaMunozWiese2022} for a detailed discussion.}
  \label{tab:flow}
\end{ruledtabular}
\end{table}

\subsection{Cole-Hopf transformation}
\label{s:Cole-Hopf}

The Cole-Hopf transformation is defined by 
\be
Z(x,t) = \rme^{\hat \lambda u(x,t)}\quad \Leftrightarrow \quad u(x,t) = \frac{\ln Z(x,t)}{\hat \lambda}. 
\ee
It is build to remove the  non-linear term proportional to $\lambda$ from the KPZ equation \eq{eq:qKPZ}, and  reproduced here, 
\bea
\eta \partial_t u(x,t) &=& c\nabla^2 u(x,t) +\lambda \left[\nabla u(x,t)  \right]^2+
m^2\big[w{-} u(x,t)\big] \nn\\
&& + F\big(x,u(x,t)\big).
\label{eq:qKPZ2}
\eea
The transformed equation reads
\bea\label{EOM-Cole-Hopf}
\eta \partial_t Z(x,t) &=& c\nabla^2 Z(x,t) +\hat \lambda Z(x,t)  F\left(x, \frac{\ln Z(x,t)}{\hat \lambda} \right) \nn\\
&& + m^2  Z(x,t) \left[\hat \lambda w-  {\ln Z(x,t)}  \right]. \qquad 
\eea
Some remarks are in order:
\begin{itemize}
\item[(i)] while the   term $\sim m^2$ in \Eq{eq:qKPZ2} provides a mass to the free propagator, i.e.\ a  decay for large distances $x$ proportional to $\rme^{-m |x|}$, it becomes a   non-linear term $\sim Z \ln Z$ in the transformed equation \eq{EOM-Cole-Hopf}. For this reason that one usually  sets $m\to 0$. 
\item[(ii)] The force $f=m^2 w$ in \Eq{eq:qKPZ2}, which could be introduced independently of the term linear in $u(x,t)$, becomes a mass for the Cole-Hopf transformed theory \eq{EOM-Cole-Hopf}, of the form $f\hat \lambda Z (x,t)$. As a result, the free propagator for $Z$ decays with a factor of  $\rme^{-|x|\sqrt{f \hat \lambda}}$.
\end{itemize}
This indicates that the Cole-Hopf transformation heavily shakes up infrared and ultraviolet properties of the theory. It may therefore not be surprising that in \PK \ no fixed point was found, whereas here, with properly defined physical fields, there is an FRG fixed point. A better understanding of the Cole-Hopf transformation and its consequences are desirable. We cannot exclude that it has some bearing on the perturbative treatment \cite{FreyTaeuber1994,Laessig1995,Wiese1997c,Wiese1998a} of the KPZ equation itself, or on the mapping between the KPZ equation and the  corresponding directed polymer problem \cite{LeDoussalWiese2005a,Wiese2021}, with all that this entails.

\section{Physical insights}\label{sec:phys_insight}

Let us summarize the main physical insights  from our work:
\begin{enumerate}
\item Most importantly,   the   qKPZ class covers a wide range of microscopic models, and is universal.
Strong evidence for this comes from the ability of the theory to predict not only the critical exponents, but also the effective KPZ amplitude $\ca A$, and the force-correlator $\Delta(w)$. 

\item The introduction of the non-linearity facilitates  depinning as compared to qEW, \Eq{Fc-main-text}.
This favors ``flatter'' interfaces, i.e.\ those for which the integrated KPZ term is smaller, reducing the roughness exponent $\zeta$. 

\item The renormalized force correlator in dimension $d=1$ is close in shape to the correlator of qEW. This means that all   properties linked to the shape of the correlator are close: for example, the avalanches-size correlations \cite{ThieryLeDoussalWiese2016}, or the correlation length $\xi_{\bot}$. The  scaling dimension  of
 $\xi_{\bot}$  is close to its qEW counterpart, $\zeta_{m}^{\text{qKPZ}} \approx \zeta^{\text{qEW}} $, whereas the roughness exponents $\zeta$ are rather different.

\item
To properly renormalize the qKPZ class, one needs a confining potential. The confining potential forbids large fluctuations of the interface, which on the technical level provides a clear distinction between short-distance and long-distance divergences. 
 \end{enumerate}

\section{Conclusion}
\label{s:Conclusions}
We revisited the   qKPZ universality class. Using a careful comparison to numerical simulations in dimensions $d=1$, $d=2$, and $d=3$, we constructed a consistent theory.  
The crucial ingredient is a flow-equation for the KPZ non-linearity, which is controlled by dimension $d$.  Behind this feature lies the observation that all    field theories for qEW with SR or LR elasticity, as well as qKPZ merge into a single theory in dimension $d=0$. Our theory has predictive powers as long as we have a sufficient knowledge of the qEW fixed point in small dimensions, and we are not too far away from $d=0$. 
We derived several bounds, respected in low dimensions, but violated in dimension $d=3$; there we currently can only close our scheme with an adhoc assumption. 

We hope that our method of first measuring the effective theory in a simulation, before attempting to build a field theory, can serve in other contexts as well. 
Applying our approach to 
other  growth experiments for which no theory is available seems promising  \cite{DiasYunkerYodhAraujoTelo-da-Gama2018}. 
We hope it will also shed light on the problems in the standard (thermal) KPZ equation in higher dimensions.

\medskip

\acknowledgements
We thank Juan A. Bonachela and Miguel A. Mu\~{n}oz for stimulating discussions and collaboration on the numerical part of this project, published in \cite{MukerjeeBonachelaMunozWiese2022}.

\appendix

\section{Field-theory details}\label{sec:diagrams}

As explained in the main text, our field theory is massive, with a time integrated response function given by $C_k=1/(c k^2+m^2)$. All diagrams 
are calculated with $C_k$. 
In appendix \ref{a:Useful momentum integrals} we first give all momentum integrals    appearing in the main text or used later.
In the following appendix \ref{a:Diagrams}, we recalculate all diagrams in the massive scheme.

\subsection{Useful momentum integrals}
\label{a:Useful momentum integrals}
To calculate all integrals, we use the Feynman representation of the time integrated response, 
\be\label{A1}
C_k = \frac1{c k^2 +m^2} = \int_{s>0} \rme^{-s (c q^2+m^2)}.
\ee
This lets appear a normalization factor 
\be\label{A2}
\int_k\rme^{-s k^2} = \frac1{(4\pi s)^{d/2}}. 
\ee
The elasticity $c$ and the mass $m$ both appear in the momentum integrals, and can be taken out by a rescaling of $k$. As an example consider 
\bea\label{a:I1}
\!\!\!\!\!I_1  &:=& \int_k \frac{1}{(c k^2 + m^2)^2} = \frac{1}{c^{d/2}} \int_k \frac{1}{(  k^2 + m^2)^2}\nn\\
& =& \frac{m^{d-4}}{c^{d/2}} \int_k \frac{1}{(k^2 + 1)^2} = \frac{m^{d-4}}{c^{d/2}}  \int_{k} \int_{s>0} s \,\rme^{- s (k^2+1)} \nn\\
&=& \frac{m^{d-4}}{c^{d/2}} \frac1{(4\pi)^{d/2}}\int_{s>0} s^{1-d/2}   =\frac{m^{d-4}}{c^{d/2}} \frac{2\Gamma(1{+}\frac \epsilon 2)}{\epsilon (4\pi)^{d/2}}.~~~~~
\eea
In the first step, we rescaled $k\to  k\sqrt c$. In the second step $k \to \frac k m$. 
These steps assume that there are no explicit cutoffs on $k$, and that the only cutoff is set by $m$, and dimensional regularization is used. 
We then    used the auxiliary integral \eq{A1}, and the momentum integral \eq{A2}.
Below we give a complete list of all encountered integrals, after rescaling to eliminate the $c$ and $m$ dependence.
\bea
\label{int100}
\int_k \frac1{k^2+1} &=& \frac{\Gamma \left(1-\frac{d}{2}\right)}{(4\pi)^{d/2}}, \\
\label{int100-bis}
\int_k \frac{k^2}{(k^2+1)^2} &=& \frac d2 \int_k \frac1{k^2+1}, \\
\label{int101}
\int_k \frac1{(k^2+1)^2} &=& \frac{\Gamma \left(2-\frac{d}{2}\right)}{(4\pi)^{d/2}}\equiv \frac{2\Gamma(1+\frac \epsilon 2)}{\epsilon (4\pi)^{d/2}},\\
\label{int102}
\int_k \frac{k_1^2}{(k^2+1)^3} &\equiv& \frac14 \int_k \frac{1}{(k^2+1)^2},  \\
\label{int103}
\int_k \frac{k^2}{(k^2+1)^3} &\equiv& \frac d4 \int_k \frac{1}{(k^2+1)^2},  \\
\label{int104}
\int_k \frac{k^4}{(k^2+1)^4} &\equiv& \frac {d(d+2)}{24} \int_k \frac{1}{(k^2+1)^2}.  \\ 
\nonumber
\eea
Integral \eq{int101} is the key-integral used to define the renormalized  force correlator $\tilde \Delta(u)$, see \Eqs{Delta-reno}-\eq{I1-main-text}. It is therefore useful to express as far as possible all integrals w.r.t.\ to integral \eq{int101}, or including the dimensions w.r.t\ integral \eq{a:I1}.

\subsection{Diagrams}
\label{a:Diagrams}

\subsubsection{The coefficient $a_0$}
According to \PK, Eq.~(A3)
\begin{eqnarray}
\diagram{KPZ1}
&=& \lambda \Delta' (0^{+})\tilde {u}\dot u \int_{k} \frac{c k^{2}}{(c 
k^{2}+m^{2})^{3}}, \qquad\qquad \label{lf95}\\
\int_{k} \frac{c k^{2}}{(c 
k^{2}+m^{2})^{3}}&=& c^{-\frac d 2} \int_{k} \frac{k^{2}}{( 
k^{2}+m^{2})^{3}}\nn\\
&=& c^{-\frac d 2}m^{d-4} \int_{k} \frac{k^{2}}{( 
k^{2}+1)^{3}}.
\end{eqnarray}
The relevant integral is \Eq{int103}, thus in \Eq{delta-eta/eta}
\be
a_0 = \frac d4.
\ee
Note that   it does not modify $\eta$ in dimension $d=0$.

\subsubsection{The coefficient $a_1$}
The first correction (in momentum space) to $\tilde u \nabla^2 u$ is
\begin{eqnarray}
\diagram{KPZ3}&=& -2 \Delta' (0^{+}) \lambda
 \int_{k}\frac{k  p}{(
c(k+p)^{2}+m^{2}) (ck^{2}+m^{2})}\nonumber \\ 
&= &  4 \Delta' (0^{+}) \lambda 
\int_{k}  \frac{
c (kp)^{2}}{(ck^{2}+m^{2})^{3}} +\ca O(p^3)\nonumber \\ 
&=&  \Delta' (0^{+}) { \hat \lambda} 
(c p^2)  I_1  \label{lf97}.
\end{eqnarray}
Note that  $-p^2 u \leftrightarrow \nabla^2 u$. This yields in \Eq{delta-c/c}
\be
a_1 = 1.
\ee

\subsubsection{The coefficient $a_2$}
The second correction (in momentum space) to $\tilde u\nabla^2  u$ is
\bea
&&\!\!\!\diagram{KPZ5} = -4 \Delta(0) \lambda^2  \int_{k}\frac{(k  p) [k(k+p)]}{(
c(k+p)^{2}+m^{2}) (ck^{2}+m^{2})^2} \nonumber \\
&& = \frac{4 \Delta(0) \lambda^2}{c^{d/2+1}}  \int_{k} \frac{2 k^2 (k  p)^2}{(k^{2}+m^{2})^4} -\frac{(k  p)^2}{(k^{2}+m^{2})^3} +\ca O(p^3)\nn\\
&& = 4 \Delta(0) \hat \lambda^2 (c p^2)   I_1    \frac{d-1}3 .  
\eea
This implies in \Eq{delta-c/c}
\be
a_2 = \frac{d-1}3.
\ee

\subsubsection{The coefficient $a_3$}
Denoting by $p_1$ and $p_2$ the momenta entering into the two external fields to the right, we have up to higher-order corrections in the $p_i$
\begin{eqnarray}
\diagram{KPZ2}&= & \int _k 4 \lambda ^2 \Delta^\prime(0^+)  \frac{(k p_1)^\alpha (k p_2)^\beta}{\left(ck^{2}+m^{2}\right)^{3}}\tilde u_{-p_1-p_2} u_{p_1} u_{p_2}\nn\\
&=& \int _k 4 \lambda ^2 \Delta^\prime(0^+)  \frac{\frac{k^2}{d}( p_1\cdot  p_2)}{\left(ck^{2}+m^{2}\right)^{3}}\tilde u_{-p_1-p_2} u_{p_1} u_{p_2}  \nonumber\\
&=&  -\int _k 4 \lambda ^2 \Delta^\prime(0^+)   \frac{\frac{k^2}{d}}{\left(ck^{2}+m^{2}\right)^{3}} \tilde u (\nabla u)^2 \nonumber\\
&=&- 4 \Delta' (0^{+}) \frac{\lambda ^2}{d}
\int_{k}  \frac{
k^{2}}{(ck^{2}+m^{2})^{3}} \tilde u (\nabla u)^2 \nonumber \\ 
&=&  -\Delta' (0^{+})   \lambda \hat \lambda
I_1 \tilde u (\nabla u)^2\label{lf97-bis}.
\end{eqnarray}
We used \Eq{int104}. This yields in \Eq{delta-lambda/lambda}
\be
a_3 = 1.
\ee

\subsubsection{The coefficient $a_4$}
\bea
\!\!\!\diagram{KPZ4} &=& -8 \frac{\Delta(0)}{d}\lambda^3 \int_{k}\frac{k^4}{{(c k^2
+ m^2 )}^4}  \tilde u (\nabla u)^2, \qquad \\
\!\!\!\diagram{KPZ6}& =& 4 \frac{\Delta(0)}{d}\lambda^3 \int_{k}\frac{k^4}{(c k^2
+ m^2 ) ^4} \tilde u (\nabla u)^2.
\eea
Together their amplitude  (without the factor of $\Delta(0)$ and $\tilde u (\nabla u)^2$) is
\be
- \frac{4 \lambda^3}{d c^{d/2+2}}   \int \frac{k^4}{(k^2+m^2)^4}  =  -\lambda   \hat \lambda ^2 \frac{d+2}{6} I_1.  
\ee
Therefore in \Eq{delta-lambda/lambda}, 
\be
a_ 4 =  \frac{d+2}{6}. 
\ee

\subsubsection{The coefficient $a_5$}
It is given by twice the integral \eq{int104}, thus for \Eq{delta-Delta/Delta}
\be
a_5 = \frac{d(d+2)}{12}.
\ee

\subsection{Depinning force}
\label{a:Fc}
The perturbative calculation gives in absence of KPZ terms
\be
F_{\rm c}^{(1)} = -\Delta'(0^+)  \int_k \frac1{c k^{2} +m^2}.
\ee
The new contribution induced by the KPZ term is 
\be
 F_{\rm c}^{(2)} = {\color{white}\fbox{\color{black}\diagram{KPZ1}}} = {\lambda}   \Delta(0)  \int_k \frac{k^2}{( c k^{2}+m^2 )^2}.
\ee
(There is a combinatorial factor of $1/2$ from $\Delta(u_t-u_t)$,  followed by a $2$ for the number of possible contractions.)
The total is 
\bea
F_{\rm c} &= &F_{\rm c}^{(1)}+F_{\rm c}^{(2)} \nn\\
&\simeq&  \left[  \Delta'(0^+) + \frac d 2  \frac\lambda c \Delta(0)\right] \int_k \frac1{c k^{2}   +m^2}.
\eea


\ifx\doi\undefined
\providecommand{\doi}[2]{\href{http://dx.doi.org/#1}{#2}}
\else
\renewcommand{\doi}[2]{\href{http://dx.doi.org/#1}{#2}}
\fi
\providecommand{\link}[2]{\href{#1}{#2}}
\providecommand{\arxiv}[1]{\href{http://arxiv.org/abs/#1}{#1}}
\providecommand{\hal}[1]{\href{https://hal.archives-ouvertes.fr/hal-#1}{hal-#1}}
\providecommand{\mrnumber}[1]{\href{https://mathscinet.ams.org/mathscinet/search/publdoc.html?pg1=MR&s1=#1&loc=fromreflist}{MR#1}}

\tableofcontents

\end{document}